\documentclass{jfm_mod}

\usepackage{graphicx, caption}
\DeclareCaptionFormat{cont}{#1 (cont.)#2#3\par}
\usepackage{color}
\usepackage[dvipsnames]{xcolor}  
\usepackage{newtxtext}
\usepackage[varvw]{newtxmath}
\usepackage{natbib}
\usepackage{hyperref}
\newcommand{\secref}[1]{\S\,\ref{#1}}
\usepackage{amsmath}
\usepackage[mathscr]{euscript}
\newcommand{\argmin}{\mathop{\mathrm{argmin\,}}}
\newcommand{\argmax}{\mathop{\mathrm{argmax\,}}}
\usepackage{multirow,multicol,makecell,hhline}
\usepackage{booktabs}
\hypersetup{
	colorlinks = true,
	urlcolor   = blue,
	citecolor  = black,
}

\newcommand{\RomanNumeralCaps}[1]
\linenumbers

\title{Prediction and control of two-dimensional decaying turbulence using generative adversarial networks}

\author{Jiyeon Kim\aff{1}, Junhyuk Kim\aff{2} \and Changhoon Lee\aff{1,3} \corresp{\email{clee@yonsei.ac.kr}}}

\affiliation{\aff{1}School of Mathematics and Computing, Yonsei University, Seoul 03722, Korea
        \aff{2}Korea Atomic Energy Research Institute, Daejeon 34057, Korea
	\aff{3}Department of Mechanical Engineering, Yonsei University, Seoul 03722, Korea}

\begin{document}
\maketitle

\begin{abstract}
With the recent rapid developments in machine learning (ML), several attempts have been made to apply ML methods to various fluid dynamics problems. However, the feasibility of ML for predicting turbulence dynamics has not yet been explored in detail. In this study, PredictionNet, a data-driven ML framework based on generative adversarial networks (GANs), was developed to predict two-dimensional (2D) decaying turbulence. The developed prediction model accurately predicted turbulent fields at a finite lead time of up to half the Eulerian integral time scale. In addition to the high accuracy in pointwise metrics, various turbulence statistics, such as the probability density function, spatial correlation function, and enstrophy spectrum, were accurately captured by the employed GAN. Scale decomposition was used to interpret the predictability depending on the spatial scale, and the role of latent variables in the discriminator network was investigated. The good performance of the GAN in predicting small-scale turbulence is attributed to the scale-selection capability of the latent variable. Results also revealed that the recursive applications of the prediction model yielded better predictions than single predictions for large lead times. Furthermore, by utilizing PredictionNet as a surrogate model, a control model named ControlNet was developed to identify disturbance fields that drive the time evolution of the flow field in the direction that optimises the specified objective function. Therefore, an illustrative example in which the evolution of 2D turbulence can be predicted within a finite time horizon and controlled using a GAN-based deep neural network is presented. 
\end{abstract}


\section{Introduction}\label{sec:intro}
Turbulence is a multiscale nonlinear dynamic phenomenon frequently observed in various flows in nature and industry. Certain deterministic dynamic features such as coherent structures have been found in turbulence~\citep{Hussain1986}; however, the behaviour of turbulence in most flows is chaotic. These characteristics make the accurate prediction of turbulence challenging despite that governing equations for turbulence, called the Navier--Stokes equations, exists. With the qualitative and quantitative expansion of computing resources over the past decades, various numerical approaches have been proposed. Direct numerical simulation (DNS), a full-resolution approach that can provide the most detailed description, is restricted to low Reynolds numbers. Moreover, the spatial filtering strategy of large eddy simulation (LES) and the temporal averaging approach of the Reynolds-averaged Navier--Stokes (RANS) model, which provide relatively fast solutions, lack reliable general closure models. Most importantly, these traditional numerical approaches are based on the temporal advancement of the governing partial differential equations, which is still costly to be used in practice, even with ever-increasing computing power.

Recently, ML- and other data-driven approaches have become popular in many areas of science, engineering, and technology owing to their effectiveness and efficiency in dealing with complex systems. Certainly, efforts have been made to apply ML to turbulence problems, particularly in fluid dynamics~\citep{Duraisamy2019, Brenner2019, brunton2020}. Other studies have attempted to develop new closure models for turbulence models using ML, such as subgrid-scale models~\citep{Gamahara2017, Beck2018, Maulik2018, Guan2021, Kim2022}. Moreover, wall models for various flows in LES have been proposed~\citep{Yang2019, Zhou2021, Bae2022, Dupuy2023, Vadrot2023, LozanoDuran2023}, and improvements in closure models for RANS have been attempted~\citep{Duraisamy2015, Ling2016, Parish2016, Singh2017, Wu2018, Duraisamy2019, Zhu2019}. Some attempts have yielded promising results; however, more exploration is required to secure reliable general closure models.

Another approach using ML to predict turbulence dynamics based on reduced-order modeling (ROM) has been proposed. With low-dimensional representations obtained through mathematical decomposition, such as proper orthogonal decomposition~\citep{Sirovich1987}, Koopman operator \citep{Mezic2005, Mezic2013} and dynamic mode decomposition~\citep{Schmid2010}, or using recent network architectures such as autoencoder (AE) and convolutional neural networks (CNNs), the governing dynamics of latent space are trained using dynamic models, such as recurrent neural networks (RNNs). For example, \citet{Hennigh2017} developed a model called Lat-Net based on the lattice Boltzmann method data by applying an AE structure. \citet{King2018} developed a compressed convolutional long short-term memory (LSTM) combining AE and LSTM and showed that dynamic approaches such as Lat-Net and their method are more effective in reflecting turbulence physics, compared to static approaches. \citet{Wang2018} and \citet{Mohan2018} efficiently predicted the coefficients of basis functions using LSTM after applying proper orthogonal decomposition to various flows. \citet{Mohan2019} extended the range of prediction to the forced homogeneous isotropic turbulence with two passive scalars. \citet{Srinivasan2019} confirmed the applicability of LSTM to wall-bounded near-wall turbulence using the nine-equation shear flow model. A recent study by \citet{Nakamura2021} successfully applied nonlinear mode decomposition to predict the minimal turbulent channel flow using a CNN-based AE. Although ROM-based methods are efficient and easy to analyze, system characteristics such as nonlinear transients and multiscale phenomena can be easily lost during information compression when only the dominant modes are used. However, as complexity of ROM-based methods increases, models tend to capture most physical phenomena of turbulence. For example, as reported by \citet{Nakamura2021}, approximately 1500 modes were found to be sufficient to fully reconstruct turbulence characteristics. Then a question arises on how many modes need to be considered and the number of modes to properly represent turbulence might be not as small as intended.


The most basic and popular ML models are artificial neural networks (ANNs), also known as multilayer perceptrons, which determine nonlinear functional relationships between the input and output data and are relatively easy to train~\citep{Beck2021}. However, when the input is in the form of a field, CNNs effectively capture embedded spatial patterns or correlations. In our previous work on the prediction of turbulent heat transfer~\citep{Kim2020b}, a CNN successfully reconstructed the wall heat-flux distribution based on wall shear stresses. However, CNN-based models whose objective function is to minimise the pointwise mean-squared difference between the predicted and target fields sometimes produce blurry outputs \citep{Kim2020, Kim2021}. Conversely, GANs~\citep{Goodfellow2014}, in which a generator ($G$) and discriminator ($D$) network are trained simultaneously in an adversarial manner such that $G$ is trained to generate high-quality data while $D$ is trained to distinguish generated data from target data, can produce better output than CNNs~\citep{Deng2019, Kim2020, Kim2021,Kim2023}. \citet{Lee2019} also showed that GANs are better in long-term prediction of unsteady flow over a circular cylinder. \citet{Ravuri2021} applied a deep generative model to precipitation nowcasting and reported a much higher accuracy with small-scale features than other ML models and numerical weather-prediction systems. GANs appear to capture the statistical characteristics of fields better than CNNs; we focus on this capability of GAN in this study.

First, we selected 2D decaying homogeneous isotropic turbulence (DHIT), which is essential in fields such as weather forecasting~\citep{Shi2015, Ruttgers2019, Liu2020, Ravuri2021}; it is relatively simple; thus, its prediction can be performed at a reasonable cost. Also, the study of 2D turbulence was initially considered as a simplified version of 3D turbulence; however, it was studied extensively \citep{Sommeria1986, Brachet1988, Mcwilliams1990, McWilliams1994, Jimenez1996} after its unique characteristics related to geophysical and astrophysical problems, such as strongly rotating stratified flow, are revealed~\citep{Alexakis2006}. The primary goal of this study was to develop a high-accuracy prediction model for 2D DHIT called PredictionNet based on GANs, that produces the evolution of turbulence by reflecting spatiotemporal statistical characteristics. Successfully trained PredictionNet could predict 2D turbulence more accurately in various aspects than a baseline CNN. Although proving why GANs are better in the prediction of turbulence statistics than CNNs is prohibitively hard, we performed various quantitative statistical analyses regarding the predictive accuracy depending on time and spatial scales to provide some clues on the working principle of the GAN model. By considering scale decomposition in the analysis of the behaviour of the latent variable, we discovered that the discriminator network of a GAN possesses a scale-selection capability, leading to the successful prediction of small-scale turbulence.

Second, flow control becomes feasible if accurate flow prediction is possible. The application of ML to turbulence control dates back to~\citet{Lee1997}, who used a neural network for turbulence control to reduce drag in a turbulent channel flow. Recently, various studies that applied ML to flow control for drag reduction in turbulent channel flow \citep{Park2020, Han2020, Lee2023}, drag reduction of flow around a cylinder \citep{Rabault2019, Rabault2019a, Tang2020}, and object control \citep{Colabrese2017, Verma2018} have been conducted, yielding successful control results. However, in this study, for a fundamental understanding of the control mechanism in 2D turbulence, we considered determining the optimum disturbance field, which can modify the flow in the direction of optimising the specified objective function. Thus, we combined PredictionNet and ControlNet for specific purposes. A target where the flow control can be used meaningfully, such as maximising the propagation of the control effect of the time-evolved flow field, was set, and the results were analyzed and compared with the results of similar studies~\citep{Jimenez2018, Yeh2021}.

Following this introduction, \secref{sec:data} describes the process of collecting datasets to be used for training and testing. In \secref{sec:machine}, ML methodologies such as objective functions and network architectures are explained. The prediction and control results are subdivided and analysed qualitatively and quantitatively in \secref{sec:results}, and a conclusion is drawn in \secref{sec:conclusion}.

\section{Data collection}\label{sec:data}
For decaying 2D turbulence, which is our target for prediction and control, DNS was performed by solving the incompressible Navier--Stokes equations in the form of the vorticity transport equation without external forcing:
\begin{equation}
	\frac{\partial \omega}{\partial t} + u_j\frac{\partial \omega}{\partial x_j} = \nu\frac{\partial^2 \omega}{\partial x_j \partial x_j}, \label{eq2.1} 
\end{equation}
with
\begin{gather}
	\frac{\partial^2 \psi}{\partial x_j \partial x_j} = -\omega, \label{eq2.2}
\end{gather}
where $\omega(x_1,x_2,t)$ is the vorticity field with $x_1=x$ and $x_2=y$ and $\nu$ is the kinematic viscosity. $\psi$ denotes the steam function that satisfies $u_1=u=\partial{\psi}/ \partial{y}$ and $u_2=v=-\partial{\psi}/\partial{x}$. A pseudo-spectral method with 3/2 zero padding was adopted for spatial discretization. The computational domain size to which the biperiodic boundary condition was applied was a square box of $[0,2\pi)^2$, and the number of spatial grids, $N_x \times N_y$, was $128\times 128$. The Crank--Nicolson method for the viscous term and second-order Adams--Bashforth method for the convective term were used for temporal advancement. In Appendix~\secref{app0}, it was proven that the pseudo-spectral approximation to Equations~(\ref{eq2.1}, \ref{eq2.2}) in a biperiodic domain is equivalent to pseudo-spectral approximation to the rotational form of the two-dimensional Navier--Stokes equations.  For the training, validation, and testing of the developed prediction network, 500, 100, and 50 independent simulations with random initialisations were performed, respectively. 

\begin{figure}
	\centerline{\includegraphics[width=0.9\columnwidth]{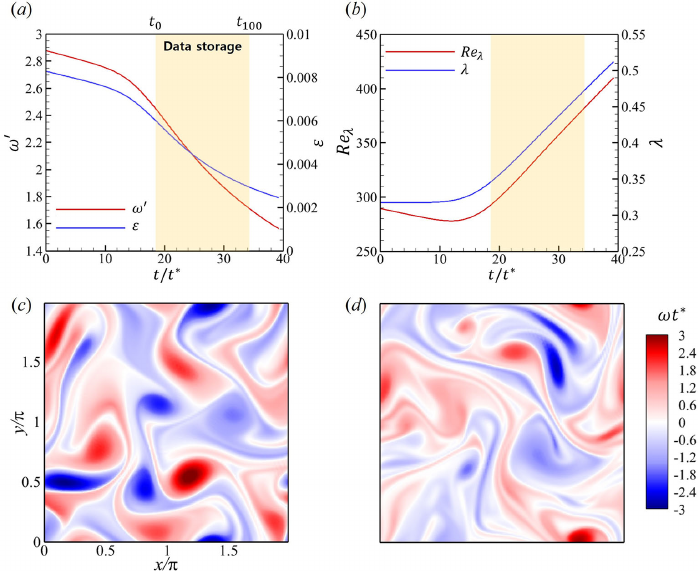}}
	\captionsetup{width=\columnwidth,justification=justified,labelsep=period}
	\caption{Range of selected data for training with a time interval of $100\delta t/t^*$ shown in the yellow region in (a) the vorticity RMS and dissipation rate and in (b) Reynolds number and the Taylor length scale. Example vorticity fields at (c) $t_0$ and (d) $t_{100}$ from a normalised test data.}
	\label{fig1}
\end{figure}

Training data were collected at discrete times, $t_i (=t_0+i \delta t,~ i=0,1,2, \cdots, 100)$ and $t_i+T$ ($T$ is the target lead time for prediction), where $\delta t (=20 \Delta t)$ is the data time step, and $\Delta t$ denotes the simulation time step. $t_0$ was selected such that the initial transient behaviour due to the prescribed initial condition in the simulation had sufficiently disappeared; thus, the power-law spectrum of enstrophy ($\Omega(k) \propto k^{-1}$ where $k$ is the wavenumber, \citealp{Brachet1988}) was maintained. During this period, the root-mean-square vorticity magnitude $\omega'$ and the average dissipation rate $\varepsilon$ decay in the form $\sim t^{-0.56}$ and $\sim t^{-1.12}$, respectively, as shown in Figure~\ref{fig1}(a), whereas the Taylor length scale ($\lambda =u'/\omega'$ where $u'$ is the RMS velocity magnitude, \citealp{Jimenez2018}) and the Reynolds number based on $\lambda$ ($Re_\lambda = u' \lambda/\nu$), which are ensemble-averaged over 500 independent simulations, linearly increase as shown in Figure \ref{fig1}(b). Therefore, 50,000 pairs of flow field snapshots were used in training, and for sufficient iterations of training without overfitting, data augmentation using a random phase shift at each iteration was adopted. Hereafter, the reference time and length scales in all nondimensionalizations are $t^*=1/\omega'_0$ and $\lambda^* =u'_0/\omega'_0$, respectively. The nondimensionalized simulation and data time steps are $\Delta t /t^*= 0.00614$ and $ \delta t/t^* = 0.123$, respectively. $t_{100}-t_0$ is approximately 2.5 times the Eulerian integral timescale of the vorticity field, as discussed below. Therefore, the vorticity fields at $t_0$ and $t_{100}$ are decorrelated as shown in Figures~\ref{fig1}(c) and \ref{fig1}(d). In our training, all of these data spanning from $t_0$ to $t_{100}$ were used without distinction such that the trained network covers diverse characteristics of decaying turbulence.

\begin{figure}
	\centerline{\includegraphics[width=0.9\columnwidth]{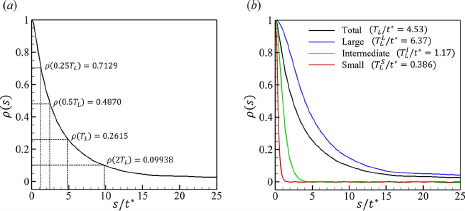}}	\captionsetup{labelsep=period}
	\caption{Distribution of the temporal autocorrelation function of (a) the whole vorticity field and (b) the scale-decomposed vorticity fields.}
	\label{fig2}
\end{figure}

To select the target lead time $T$, we investigate the temporal autocorrelation function of the vorticity field, $\rho(s) (= \left< \omega(t)\omega(t+s) \right> / \left< \omega(t)^2 \right>^{1/2} \left< \omega(t+s)^2 \right>^{1/2})$ for time lag $s$, as shown in Figure \ref{fig2}(a), from which the integral time scale is obtained, $T_L = 4.53 t^* = 36.9 \delta t$. As it is meaningless to predict the flow field much later than one integral time scale, we selected four lead times to develop a prediction network: $10\delta t, 20 \delta t, 40 \delta t$, and $80 \delta t$, which are referred to as $0.25 T_L, 0.5T_L, T_L$, and $2T_L$, respectively, even though $T_L = 36.9 \delta t$. Figure \ref{fig2}(b) shows the correlation function for the scale-decomposed field of vorticity, where three decomposed fields are considered: a large-scale field consisting of the wavenumber components for $k \leq 4$, representing the energy (enstrophy)-containing range; an intermediate-scale field containing the inertial-range wavenumber components for $4<k \leq 20$; and the small-scale field corresponding to the wavenumber components for $k>20$ in the dissipation range. This clearly illustrates that the large-scale field persists longer than the total field with an integral time scale $T^L_L \simeq 1.4 T_L$, whereas the intermediate- and small-scale fields quickly decorrelate with $T_L^I \simeq 0.25 T_L$ and $T_L^S \simeq 0.09 T_L$. These behaviours are responsible for the different prediction capabilities of each scale component, as discussed later.

\begin{figure}
	\centerline{\includegraphics[width=0.45\columnwidth]{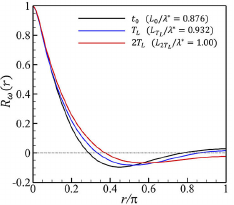}}	\captionsetup{labelsep=period}
	\caption{Ensemble averaged two-point correlation functions of vorticity extracted from 500 training data.}
	\label{fig3}
\end{figure}

The spatial two-point correlation function of vorticity $R_{\omega}(r,t)$ with the corresponding integral length scale $L_t$ at three different times $t=t_0, t_0+T_L$, and $t_0+2T_L$ is shown in Figure \ref{fig3}. For the investigated time range, $R_{\omega}(r,t)$ decays sufficiently close to zero at $r=\pi$ (half domain length), even though $L_t$ tends to increase over time from $0.876\lambda^*$ at $t_0$ to $1.03\lambda^*$ at $t_{100}$ because of the suppression of the small-scale motions of 2D turbulence. This marginally decaying nature in the periodic domain was considered in the design of the training network such that data at all grid points were used in the prediction of vorticity at one point, as discussed in the following section.

\section{Machine learning methodology}\label{sec:machine}
\subsection{ML models and objective functions}\label{sub:overall}
Training ANNs is the process of updating weight parameters to satisfy the nonlinear relationships between the inputs and outputs as closely as possible. The weight parameters were optimised to minimise the prescribed loss function (in the direction opposite to the gradient) by reflecting nonlinear mappings. Loss functions are mainly determined by objective functions, and other loss terms are often added to improve training efficiency and model performance. In GANs, a generator ($G$) and discriminator ($D$) are trained simultaneously in an adversarial manner; parameters of $G$ and $D$ are iteratively and alternately updated to minimize $\log(1-D(G(\boldsymbol{z})))$ for $G$ and maximize $\log(D(\boldsymbol{x}))+\log(1-D(G(\boldsymbol{z})))$ for $D$, respectively. This stands for the two-player min-max game with a value function $V(G,D)$ given by
\begin{equation}\label{eq3.1}
	\underset{G}{\mathrm{min}}\,\underset{D}{\mathrm{max}}V(G,D) = \mathbb{E}_{\boldsymbol{x}\sim p(\boldsymbol{x})} \left[\log D(\boldsymbol{x})\right] + \mathbb{E}_{\boldsymbol{z}\sim p(\boldsymbol{z})} \left[\log (1-D(G(\boldsymbol{z})))\right],
\end{equation}
where $\boldsymbol{x}$ and $\boldsymbol{z}$ are real data and random noise vectors, respectively. Operator $\mathbb{E}$ denotes the expectation over some sampled data, and the expressions $\boldsymbol{x}\sim p(\boldsymbol{x})$ and $\boldsymbol{z}\sim p(\boldsymbol{z})$ indicate that $\boldsymbol{x}$ is sampled from the distribution of the real dataset $p(\boldsymbol{x})$ and $\boldsymbol{z}$ from some simple noise distribution $p(\boldsymbol{z})$ such as a Gaussian distribution, respectively. Thus, we can obtain a generator that produces more realistic images. Various GANs have been developed rapidly since their introduction \citep{Mirza2014, Arjovsky2017, Gulrajani2017, Karras2017, Mescheder2018, Park2019, Zhu2020}. Among these, a conditional GAN (cGAN)~\citep{Mirza2014} provides additional information $\boldsymbol{y}$, which can be any type of auxiliary information, as a condition for the input of the generator and discriminator to improve the output quality of the generator, as follows:
\begin{equation}\label{eq3.2}
	\underset{G}{\mathrm{min}}\,\underset{D}{\mathrm{max}}V(G,D) = \mathbb{E}_{\boldsymbol{x}\sim p(\boldsymbol{x})} \left[\log D(\boldsymbol{x\mid \boldsymbol{y}})\right] + \mathbb{E}_{\boldsymbol{z}\sim p(\boldsymbol{z})} \left[\log (1-D(G(\boldsymbol{z\mid \boldsymbol{y}})))\right].
\end{equation}
Furthermore, we employ two adaptive methods that can stabilise the training process, solve the problem of the vanishing gradient in which the discriminator is saturated, and prevent mode collapse, a phenomenon in which the distribution of generated samples is restricted to a specific small domain, even though the generator does not diverge. First, Equation~(\ref{eq3.2}) is modified using the Earth-Mover (EM) (Wasserstein-1) distance combined with the Kantorovich-Rubinstein (KR) duality~\citep{Villani2009}, which is called Wasserstein-GAN (WGAN)~\citep{Arjovsky2017}, as shown in Equation~(\ref{eq3.3}).
\begin{equation}\label{eq3.3}
	\underset{G}{\mathrm{min}}\,\underset{D}{\mathrm{max}}\,V(G,D) = \mathbb{E}_{\boldsymbol{x} \sim p(\boldsymbol{x})} \left[ D(\boldsymbol{x \mid \boldsymbol{y}}) \right] - \mathbb{E}_{\boldsymbol{z} \sim p(\boldsymbol{z})} \left[ D(G(\boldsymbol{z \mid \boldsymbol{y}})) \right].
\end{equation}
This modification is made based on thorough examinations of various ways of measuring the distance between the real (data) distribution ($p_r$) and the model distribution  ($p_g$), including the total variation distance, Kullback-Leibler divergence, and Jensen-Shannon divergence. EM distance can be expressed as the final form of Equation~(\ref{eq3.4}) by the KR duality:
\begin{equation}\label{eq3.4}
	\mathrm{EM}(p_r,p_g) = \underset{\| f \|_{L} \leq 1}{\sup} \mathbb{E}_{\boldsymbol{x} \sim p_r} \left[ f(\boldsymbol{x}) \right]-\mathbb{E}_{\boldsymbol{x} \sim p_g} \left[ f(\boldsymbol{x}) \right],
\end{equation}
where the supremum is taken over all the 1-Lipschitz functions for the set of real data $\mathcal{X}$, $f : \mathcal{X} \to \mathbb{R}$. Simply put, the model $g$ that wants to make $p_g$ close to $p_r$ represents the generator, and $f$ corresponds to the discriminator that is optimized to make the distance between  $p_r$ and $p_g$ larger. Thus, it can be melted down to the form of Equation~(\ref{eq3.3}). Then, a gradient penalty (GP) loss term is added to obtain the final form of WGAN-GP \citep{Gulrajani2017}. We intend to develop a model capable of predicting the dynamic behaviour of turbulence with high predictive accuracy by reflecting statistical aspects, employing a cGAN with WGAN-GP for PredictionNet and comparing the results with a baseline CNN.

\begin{figure}
	\centerline{\includegraphics[width=0.95\columnwidth]{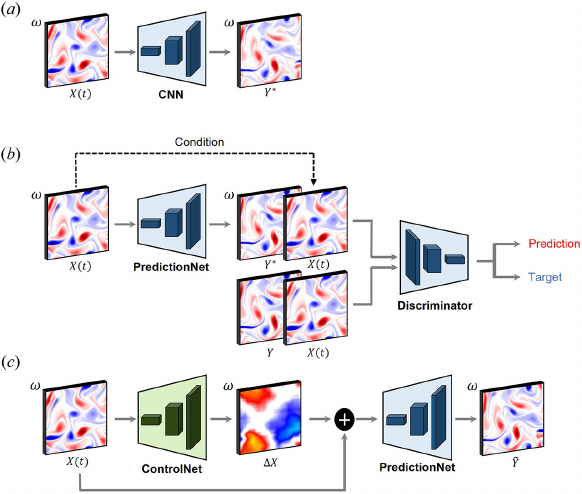}}
	\captionsetup{labelsep=period}
	\caption{Simplified network schematics of (a) the baseline CNN, (b) cGAN-based PredictionNet, and (c) ControlNet.}
	\label{fig4}
\end{figure}

PredictionNet is a network that predicts the vorticity field after a specific lead time $T$ from the input field at $t$ as follows:
\begin{equation}\label{eq3.5}
	Pred(X(t))={Y^*}\approx X(t+T)=Y,
\end{equation}
where $X$, $Y^*$, and $Y$ represent the real data, prediction results, and prediction targets, respectively (here, the prediction target $Y$ is independent of the additional input $\boldsymbol{y}$ in Equations (\ref{eq3.2}) and (\ref{eq3.3}), which are general descriptions of the value function of each GAN model). The prediction network is trained using DNS data to play a functional role in predicting the vorticity field after the lead time from each time in our dataset. Therefore, the following objective function becomes the optimisation target, regardless of the applied model:
\begin{equation}\label{eq3.6}
	\underset{w_p}{\argmin}\|{Y^*- Y}\|,
\end{equation}
with the trainable parameters of PredictionNet $w_p$, and $\|\cdot\|$ represents a distance norm of any type. PredictionNet is the generator when the GAN algorithm is applied and the adversarial loss term is added to the objective function above. In our cGAN application, the generator input ($X$), which was used to generate the output ($Y^*$), is also used as a condition in the training of the discriminator. This allows the statistical characteristics of the input ($X$) as well as the target ($Y$) to be reflected on the training of the discriminator and eventually the generator through competitive training. Conditioning is implemented through concatenation so that both the generator output $Y^*$ and $X$ or the target $Y$ and $X$ are used as input to the discriminator as illustrated in Figure~\ref{fig4}(b).  The final form of the loss function of cGAN, including the GP term, is as follows:
\begin{gather}\label{eq3.7}
	L_{pred}=\gamma\mathbb{E}_{X\sim p(X)} \left[\|{Y^*- Y}\|_2^2\right]-L_{adv}, \quad
	L_{adv}=L_{false},
\end{gather}
for the generator and
\begin{gather}\label{eq3.8}
	L_{D}=-L_{true}+L_{false}+\alpha L_{gp}+\beta L_{drift}, 
\end{gather}
for the discriminator, where
\begin{gather}\label{eq3.9}
	L_{true}=\mathbb{E}_{X\sim p(X)} \left[D(Y,X)\right], \quad
	L_{false}=\mathbb{E}_{X\sim p(X)} \left[D(Y^*,X)\right], \nonumber\\[3pt]
	L_{gp}=\mathbb{E}_{X'\sim p(X')} \left[\left(\left\|\nabla_{X'}D(X') \right\|_2-1\right)^2 \right],
	L_{drift}=\left(\mathbb{E}_{X\sim p(X)} \left[D(Y,X)\right]\right)^2,
\end{gather}
where $X'=Y+\delta(Y^*-Y)$ with $\delta$ between zero and one. The simplest L2 norm distance was used for data loss. The role of $L_{drift}$ was to restrict the order of discriminator outputs (keeping them from drifting too far from zero) with a small weight $\beta$. Its original form in \citet{Karras2017} is similar to the L2 regularization on $D(Y,X)$ as $L_{drift}=\mathbb{E}_{X\sim p(X)} \left[\left(D(Y,X)\right)^2\right]$, but we modified it to the above form, in which regularization can be applied average-wise during the backpropagation to robustly use hyperparameters $\alpha$ and $\beta$ regardless of the lead time. Accordingly, $\beta=0.001$ was used equally for all lead times, and $\alpha$ and $\gamma$ were fixed at 10 and $100/(N_x \times N_y)$, respectively, by fine tuning. There exists a separate optimum hyperparameter setting for each lead time; however, we verified that our hyperparameter setting showed no significant difference in performance from the optimum settings. In addition, we verified that it worked properly for lead times ranging from $1\delta t$ to $100\delta t$. For the loss function of the baseline CNN, L2 regularisation loss was added to Equation (\ref{eq3.6}) using L2 norm distance to improve the performance and make the optimisation process efficient, as follows:
\begin{gather}\label{eq3.10}
	L_{CNN}=\sigma_1 \mathbb{E}_{X\sim p(X)} \left[\|{Y^*- Y}\|_2^2\right]+\sigma_2 R(w_p), \quad R(w_p)=\frac12\|w_{p}\|_2^2,
\end{gather}
where $\sigma_1$ is set to $1/(N_x \times N_y)$, and $\sigma_2$ modulates the strength of regularisation and is fixed at $0.0001$ based on case studies. Simplified configurations of the CNN and cGAN are shown in Figure~\ref{fig4}(a) and \ref{fig4}(b). 

ControlNet uses a pre-trained PredictionNet that contains the dynamics of our data as a surrogate model to generate disturbance fields that change the evolution of the flow field in a direction suitable for a target, as follows:
\begin{gather}\label{eq3.11}
	Control(X(t))=\Delta X, \quad
	Pred(X(t)+\Delta X)=\tilde{Y}.
\end{gather}
$\Delta X$ represents the disturbance field added to the input field $X(t)$, and $\tilde{Y}$ is the disturbance-added prediction at time $t+T$. An important implication of ControlNet in this study is that it is a model-free method without restrictions, except for the strength of $\Delta X$. The objective function to be maximized includes  the change in the vorticity field at a later time, as follows: 
\begin{gather}\label{eq3.12}
	\underset{w_c}{\argmax}\|{Y^*- \tilde{Y}}\|, 
\end{gather}
where $w_c$ are the weight parameters of ControlNet. In the process of training of ControlNet, the weight parameters of ControlNet are updated in the direction maximizing the change in the vorticity field at the target lead time. The trained PredictionNet with fixed weight parameters is used in the prediction of controlled field as a surrogate model in training of ControlNet. Therefore, once the training of ControlNet is completed through maximization of the loss based on the change in the vorticity field, the trained ControlNet can produce an optimum disturbance field. Whether the generated disturbance field is globally optimum, however, is not guaranteed. For the final form of the loss function, a spectral gradient loss term is additionally used to remove nonphysical noise (i.e. smoothing effect), and a CNN model is applied:
\begin{gather}\label{eq3.13}
	L_{control}=\mathbb{E}_{X\sim p(X)} \left[\|{Y^*- \tilde{Y}}\|_2^2\right]+\theta L_{grad}, \quad
	L_{grad}=\mathbb{E}_{X\sim p(X)} \left[\|\nabla_{\boldsymbol{x}}\left(\Delta X\right)\|_2^2\right],
\end{gather}
where $\nabla_{\boldsymbol{x}}$ denotes the gradient with respect to the coordinate directions. The coefficient $\theta$ controls the strength of smoothing effect, and it is adjusted depending on the order of data loss (see \secref{sub:control1}). Figure \ref{fig4}(c) shows a simplified configuration of ControlNet.

\subsection{Network architecture}\label{sub:archi}
\begin{figure}
	\centerline{\includegraphics[width=1.0\columnwidth]{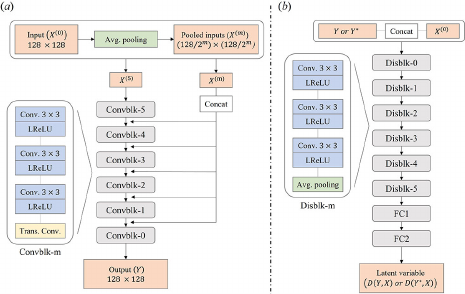}}
	\captionsetup{width=\columnwidth,justification=justified,labelsep=period}
	\caption{Network architecture of (a) PredictionNet (the generator of cGAN) and ControlNet and (b) discriminator of cGAN. }
	\label{fig5}
\end{figure}

A typical multiscale architecture of a CNN was used for both PredictionNet and ControlNet, as shown in Figure~\ref{fig5}(a). The architecture consists of six convolutional blocks, each composed of three convolutional layers with $3 \times 3$ filter kernels (Conv. $3 \times 3$), called Convblk-m, named after their feature maps with different resolutions (${128/2^m}\times{128/2^m} \; (m=0,1,2,3,4,5)$). Here, the average pooled inputs $X^{(m)} \;(m=1,2,3,4)$ and the original input $X^{(0)}$ are concatenated to the feature map tensor at the beginning of each Convblk-m to be used as feature maps. One node point of a pooled input $X^{(m)}$ contains the compressed information of $2^m$ points of $X^{(0)}$. Using such an architecture enables all the spatial grid information of an input field, $X$, to be used to calculate a specific output point, even for the lowest-resolution feature maps. For example, the receptive field of Convblk-5 is $(2^5\times7)\times (2^5\times7)$, which means that all the spatial grid points of $X^{(0)}$ are used to calculate a point in the output. In addition, by sequentially using low- to high-resolution feature maps, the network can learn large-scale motion in the early stages, and then fill in the detailed features of small-scale characteristics in the deepest layers. As mentioned in \secref{sec:data}, the purpose of designing networks is to increase the prediction accuracy and determine the global optimum by considering the spectral properties of the flow. However, depending on the flow features or characteristics of the flow variables, we can expect that accurate results can be obtained even using only a small input domain with a size similar to $L$ for cost-effectiveness. Furthermore, periodic padding was used to maintain the size of the feature maps after convolutional operations, because the biperiodic boundary condition was applied spatially.
The feature maps generated from Convblk-(m+1) must be extended through upscaling to be concatenated and used with the following pooled input $X^{(m)}$ (i.e. doubling the size via upsampling). In this process, a transposed convolution is used to minimise the nonphysical phenomena. After upscaling, $X^{(m)}$ is concatenated to the feature map tensor and then Convblk-m operations are performed. Finally, after the original input field $X^{(0)}$ is concatenated, the operations of Convblk-0 are performed, and the output of resolution ${128}\times{128}$ is generated depending on the network, prediction result $Pred(X)=Y^*$, or disturbance field $Control(X)=\Delta X$. Detailed information on the architectures of PredictionNet and ControlNet, including the number of feature maps used in each Convblk, are presented in Table~\ref{table1}. A leaky rectified linear unit (LReLU) is used as an activation function to add nonlinearities after each convolutional layer. This was not applied to the last layer of Convblk-0, to prevent nonlinearity in the final output.

The discriminator of PredictionNet to which cGAN is applied has a symmetric architecture for the generator (PredictionNet), as shown in Figure~\ref{fig5}(b). $X^{0}$ is concatenated with the target or the prediction result for the input of the discriminator as a condition. In contrast to the generator, convolutional operations are performed from high to low resolutions, named with each convolutional block as Disblk-m. Average pooling was used for downsampling to half the size of the feature maps. After the operation of Disblk-5, its output feature maps passed through two fully connected layers (with output dimensions of $1\times 1\times 256$ and $1\times 1\times 1$) to return a scalar value. The numbers of feature maps used for each Disblk are presented in Table~\ref{table1}. The baseline CNN model takes the same architecture as PredictionNet but is trained without adversarial training through the discriminator.
\begin{table}\centering
	\begin{tabular}{c c c | c c c}
		Convblk-m \;& Resolution \;& $\#$ of feature maps \quad&\;\quad Disblk-m \;& Resolution \;& $\#$ of feature maps \\
		\hline
		Convblk-5 \;& $4\times 4$ \;& 64, 64, 64 \quad&\;\quad Disblk-0 \;& $128\times 128$ \;& 16, 16, 32 \\[2pt]
		Convblk-4 \;& $8\times 8$ \;& 64, 64, 64 \quad&\;\quad Disblk-1 \;& $64\times 64$ \;& 32, 32, 64 \\[2pt]
		Convblk-3 \;& $16\times 16$ \;& 64, 64, 64 \quad&\;\quad Disblk-2 \;& $32\times 32$ \;& 64, 64, 64 \\[2pt]
		Convblk-2 \;& $32\times 32$ \;& 64, 64, 64 \quad&\;\quad Disblk-3 \;& $16\times 16$ \;& 64, 64, 64 \\[2pt]
		Convblk-1 \;& $64\times 64$ \;& 32, 32, 32 \quad&\;\quad Disblk-4 \;& $8\times 8$ \;& 64, 64, 64 \\[2pt]
		Convblk-0 \;& $128\times 128$ \;& 16, 16, 1 \quad&\;\quad Disblk-5 \;& $4\times 4$ \;& 64, 64, 64 \\[2pt]
		\hline
	\end{tabular}
	\captionsetup{width=\columnwidth,labelsep=period}
	\caption{Number of feature maps used at each convolutional layer of Convblks and Disblks.}
	\label{table1}
\end{table}

\section{Results}\label{sec:results}
\subsection{PredictionNet – prediction of the vorticity field at a finite lead time}\label{sub:pred1}
\begin{figure}
	\centerline{\includegraphics[width=0.9\columnwidth]{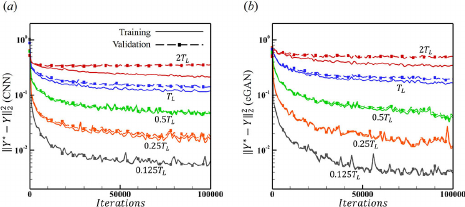}}
	\captionsetup{width=\columnwidth,justification=justified,labelsep=period}
	\caption{Convergence of data loss depending on the lead time of (a) baseline CNN and (b) cGAN. The order of the converged value increases as the lead time gets larger. Compared with the range of normalised vorticity $X$ at $t_0$, $(-3, 3)$, $T=2T_L$ has a relatively large value of converged data loss (approximately 0.25 and 0.4 for CNN and cGAN, respectively, which are approximately 8.3\% and 13.3\% of the maximum value). In addition, relatively large overfitting was observed in the case of $T=2T_L$.}
	\label{fig6}
\end{figure}

In this section, we discuss the performance of PredictionNet in predicting the target vorticity field at $t+T$ using the field at $t$ as the input. The considered lead times $T$ are $0.25T_L, 0.5T_L, T_L$, and $2T_L$ based on the observation that the autocorrelation of the vorticity at each lead time dropped to 0.71, 0.49, 0.26, and 0.10, respectively (Figure~\ref{fig2}(a)). The convergence behaviour of the L2 norm-based data loss of the baseline CNN and cGAN models for various lead times in the process of training is shown for 100,000 iterations with a batch size of 32 in Figure~\ref{fig6}. It took around 2 and 5.4 hours for CNN and cGAN, respectively, on a GPU machine of NVIDIA GeForce RTX 3090. Although the baseline CNN and PredictionNet (generator of cGAN) have the exact same architecture (around 520k trainable parameters based on Table~\ref{table1}), cGAN includes a discriminator to be trained comparable to the generator with many complex loss terms; thus, the training time is almost tripled. Both models yielded good convergence for lead times of up to $0.5T_L$, whereas overfitting was observed for $T=T_L$ and $2T_L$ in both models. Since the flow field is hardly correlated with the input field (correlation coefficient of 0.26 and 0.10), it is naturally more challenging to train the model that can reduce an MSE-based data loss, a pointwise metric that is independent of the spatial distribution. One notable observation in Figure~\ref{fig6} is that for lead times beyond $T_L$, CNN appears to exhibit better performance than cGAN, as evidenced by its smaller converged data loss. However, as will be discussed later, the pointwise accuracy alone cannot cover all the aspects of turbulence prediction due to its various statistical properties. Additionally, although the only loss term of CNN is the pointwise MSE except the weight regularization, the magnitude of its converged data loss also remains significant.

\begin{figure}
	\centerline{\includegraphics[width=0.95\columnwidth]{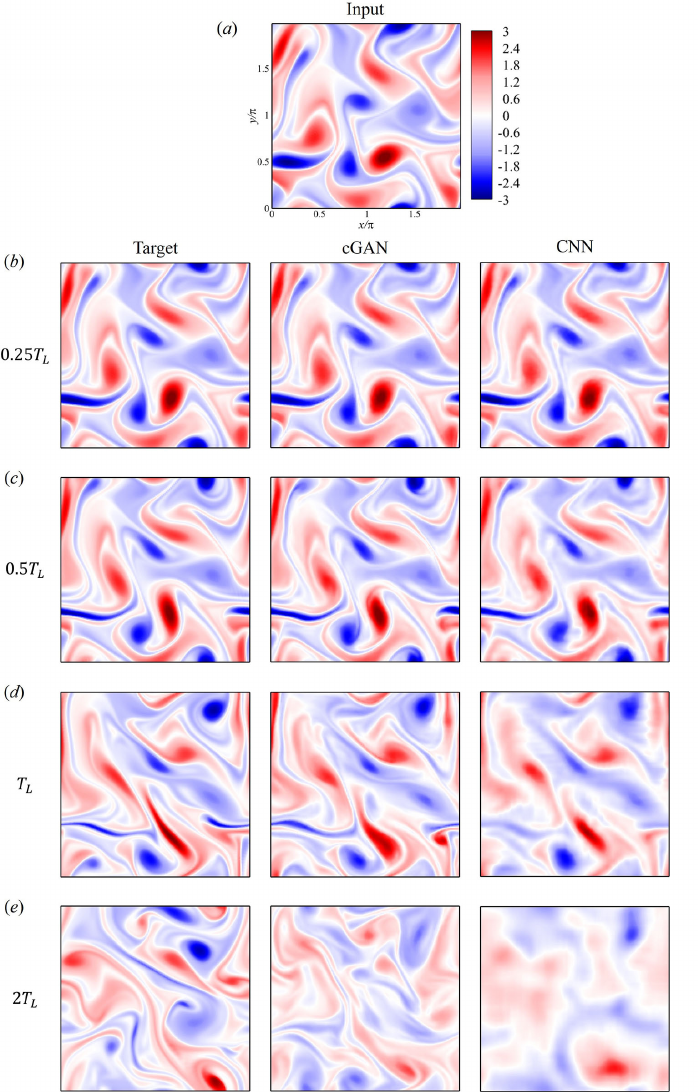}}
	\captionsetup{width=\columnwidth,justification=justified,labelsep=period}
	\caption{Visualised examples of the performance of the cGAN and CNN for one test dataset. (a) Input field at $t_0$, prediction results at the lead time (b) $0.25T_L$, (c) $0.5 T_L$, (d) $T_L$, and (e) $2T_L$. The first, second, and third columns show the target DNS, cGAN predictions, and CNN predictions, respectively.}
	\label{fig7}
\end{figure}

As an example of the test of the trained network, for unseen input data, the fields predicted by cGAN and CNN were compared against the target data for various lead times, as shown in Figure~\ref{fig7}. In the test of cGAN, only PredictionNet was used to generate or predict flow field at the target lead time. Hereinafter, when presenting performance results for the instantaneous field, including Figure~\ref{fig7} and all subsequent statistics, we only brought the results for input at $t_0$. This choice was made because the difficulty of tasks that the models have to learn is much larger at earlier time since flow contains more small-scale structures than later times due to decaying nature. Therefore, performance test for input data at $t_0$ is sufficient for test of the trained network.  For short lead-time predictions, such as $0.25T_L$, both cGAN and CNN showed good performance. However, for a lead time of $0.5T_L$, the prediction by the CNN started displaying a blurry image, whereas cGAN's prediction maintained small-scale features relatively better than CNN. This blurry nature of the CNN worsened the predictions of $T_L$ and $2T_L$. Although cGAN's prediction also deteriorated as the lead time increased, the overall performance of cGAN, particularly in capturing small-scale variations in the field, was much better than that of the CNN even for $T_L$ and $2T_L$.

A quantitative comparison of the performances of the cGAN and CNN against the target in the prediction of various statistics is presented in Table~\ref{table2}. The correlation coefficient (CC) between the prediction and target fields or, equivalently, the mean-squared error by the cGAN shows a better performance than that by the CNN for lead times of up to $0.5T_L$, even though both the cGAN and CNN predicted the target field well. For $T_L$, the CC by cGAN and CNN was 0.855 and 0.887, respectively, indicating that the predictions by both models were not poor, whereas the predictions by both models were quite poor for $2T_L$. Once again, for lead times larger than $T_L$, CNN shows better performance on the pointwise metrics according to the trend of data loss. Conversely, the statistics related to the distribution of vorticity, such as the RMS value or standardised moments ($\hat{\mu}_n=\mu_n/\sigma^n$, where $\mu_n=\left<(X-\left<X\right>)^n\right>$ is the $n^{th}$ central moments and $\sigma=\sqrt{\mu_2}$ is the standard deviation), clearly confirm that the cGAN outperforms the CNN for all lead times, and the prediction by the cGAN was much closer to the target than that of the CNN, even for $T_L$ and $2T_L$. Furthermore, the prediction of the Kolmogorov length scale ($\eta=\left(\nu^3/\varepsilon\right)^{1/4}$) and dissipation ($\varepsilon=2\nu\left<S_{ij}S_{ij}\right>$, $\varepsilon'=\varepsilon t^{*3}/\lambda^{*2}$, where $S_{ij}$ is the strain rate tensor) by cGAN was more accurate than that by the CNN, as shown in Table~\ref{table2}. All these qualitative and quantitative comparisons clearly suggest that the adversarial learning of the cGAN tends to capture the characteristics of turbulence data better than the CNN, which minimises the pointwise MSE only. The superiority of the cGAN is more pronounced for large lead times, for which small-scale turbulence features are difficult to predict.

\begin{table}\centering
	\begin{tabular}{c c c c | c c c c c | c c}
		\makecell[c]{lead time $T$ \\$(\rho(T))$} & \;& CC & MSE \;&\; RMS & $\hat{\mu}_4$ & $\hat{\mu}_6$ & $\hat{\mu}_8$ & $\hat{\mu}_{10}$ \;&\; $\eta/\lambda^*$ & $\varepsilon' (\times 10^{-2})$ \\
		\hline
		\multirow{3}{*}{\makecell[c]{$0.125T_L$\\(0.8774)}} & Target \;& - & - \;&\; 0.9871 & 2.963 & 13.70 & 80.34 & 543.3 \;&\; 0.1443 & 2.001 \\[1pt]
		& cGAN \;& 0.9983 & 3.32e-3 \;&\; 0.9813 & 2.957 & 13.64 & 79.82 & 539.0 \;&\; 0.1446 & 1.978 \\[1pt]
		& CNN \;& 0.9968 & 6.05e-3 \;&\; 0.9813 & 2.961 & 13.67 & 79.95 & 539.1 \;&\; 0.1447 & 1.978 \\
		\hline
		\multirow{3}{*}{\makecell[c]{$0.25T_L$\\(0.7129)}} & Target \;& - & - \;&\; 0.9728 & 3.001 & 14.15 & 84.87 & 587.8 \;&\; 0.1453 & 1.945 \\[1pt]
		& cGAN \;& 0.9942 & 1.09e-2 \;&\; 0.9643 & 3.016 & 14.36 & 87.29 & 614.7 \;&\; 0.1459 & 1.911 \\[1pt]
		& CNN \;& 0.9900 & 1.85e-2 \;&\; 0.9603 & 3.019 & 14.42 & 88.25 & 627.6 \;&\; 0.1463 & 1.895 \\
		\hline
		\multirow{3}{*}{\makecell[c]{$0.5T_L$\\(0.4870)}} & Target \;& - & - \;&\; 0.9440 & 3.084 & 15.14 & 95.05 & 690.5 \;&\; 0.1475 & 1.832 \\[1pt]
		& cGAN \;& 0.9692 & 0.0537 \;&\; 0.9357 & 3.108 & 15.56 & 101.2 & 776.2 \;&\; 0.1482 & 1.800 \\[1pt]
		& CNN \;& 0.9681 & 0.0557 \;&\; 0.8917 & 3.152 & 16.05 & 105.5 & 806.6 \;&\; 0.1518 & 1.636 \\
		\hline
		\multirow{3}{*}{\makecell[c]{$T_L$\\(0.2615)}} & Target \;& - & - \;&\; 0.8877 & 3.268 & 17.46 & 120.6 & 966.9 \;&\; 0.1522 & 1.621 \\[1pt]
		& cGAN \;& 0.8293 & 0.262 \;&\; 0.8603 & 3.261 & 17.40 & 121.5 & 1010 \;&\; 0.1546 & 1.523 \\[1pt]
		& CNN \;& 0.8676 & 0.196 \;&\; 0.7946 & 3.413 & 19.49 & 146.8 & 1307 \;&\; 0.1610 & 1.301 \\
		\hline
		\multirow{3}{*}{\makecell[c]{$2T_L$\\(0.09938)}} & Target \;& - & - \;&\; 0.7883 & 3.648 & 22.90 & 189.1 & 1827 \;&\; 0.1615 & 1.279 \\[1pt]
		& cGAN \;& 0.3693 & 0.708 \;&\; 0.6976 & 3.569 & 22.15 & 189.2 & 2000 \;&\; 0.1717 & 1.002 \\[1pt]
		& CNN \;& 0.4389 & 0.562 \;&\; 0.5795 & 4.064 & 29.73 & 303.4 & 3865 \;&\;0.1889 & 0.696 \\ 
		\hline
	\end{tabular}
	\captionsetup{labelsep=period}
	\caption{Quantitative comparison of the performance of the cGAN and CNN against the target in the prediction of the CC and MSE \& RMS and the $n^{th}$ standardized moments \& Kolmogorov length scale ($\eta$) and dissipation rate ($\varepsilon'$) depending on the lead time for the input at $t_0$. All the dimensional data such as MSE, RMS, and $\varepsilon'$ are provided just for the relative comparison.}
	\label{table2}
\end{table}

\begin{figure}
	\centerline{\includegraphics[width=1.0\columnwidth]{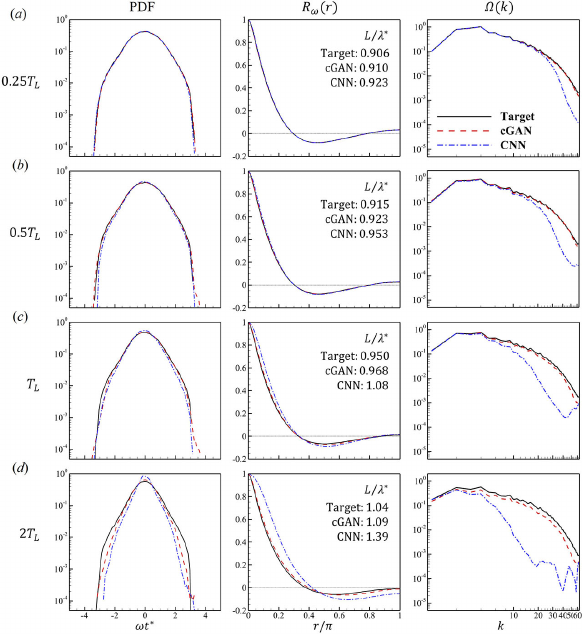}}
	\captionsetup{width=\columnwidth,justification=justified,labelsep=period}
	\caption{Comparison of statistics such as the PDF, two-point correlation, and enstrophy spectrum of the target and prediction results at different lead times (a) $0.25T_L$, (b) $0.5 T_L$, (c) $T_L$, and (d) $2T_L$ for the same input time point $t_0$. }
	\label{fig8}
\end{figure}

Detailed distributions of probability density functions (PDFs) of vorticity, two-point correlation functions, and enstrophy spectra ($\Omega(k)=\pi k\Sigma_{|\boldsymbol{k}|=k}\left|\hat{\omega}(\boldsymbol{k})\right|^2$ where $\hat{\omega}(\boldsymbol{k},t)=\mathcal{F}\left\{\omega(\boldsymbol{x},t)\right\}$) of the target and prediction fields by the cGAN and CNN are compared for all lead times in Figure~\ref{fig8}. Both the cGAN and CNN effectively predicted the target PDF distribution for lead times of up to $T_L$, whereas the tail part of the PDF for $2T_L$ was not effectively captured by the cGAN and CNN. The difference in performance between the cGAN and CNN is more striking in the prediction of $R_{\omega}(r)$ and $\Omega(k)$. $R_{\omega}(r)$ and $\Omega(k)$ obtained by the cGAN almost perfectly match those of the target for all lead times, whereas those predicted by the CNN show large deviations from those of the target distribution progressively with the lead time. In the prediction of the correlation function, the CNN tends to produce a relatively slow decaying distribution compared with the cGAN and the target, resulting in a much larger integral length scale for all lead times. In particular, it is noticeable that even for a short lead time of $0.25T_L$, the CNN significantly underpredicts $\Omega(k)$ in the high-wavenumber range, and this poor predictability propagates toward the small-wavenumber range as the lead time increases, whereas cGAN produces excellent predictions over all scale ranges, even for $2T_L$. This evidence supports the idea that adversarial learning accurately captures the small-scale statistical features of turbulence.

\begin{figure}
	\centerline{\includegraphics[width=1.0\columnwidth]{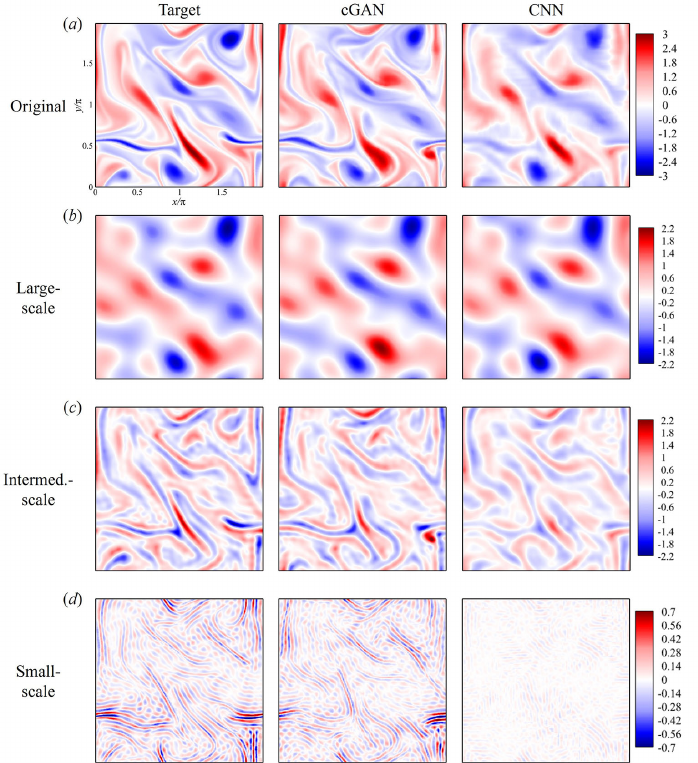}}
	\captionsetup{width=\columnwidth,justification=justified,labelsep=period}
	\caption{Prediction results of the scale-decomposed field for lead time $T_L$. (a) Original fields, (b) large scales ($k\leq 4$), (c) intermediate scales ($4< k \leq 20$), and (d) small scales ($k>20$). The first, second, and third columns display the target DNS fields, cGAN predictions, and CNN predictions, respectively.}
	\label{fig9}
\end{figure}

To quantify in more detail the differences in the prediction results by cGAN and CNN models, scale decomposition was performed by decomposing the fields in the wavenumber components into three regimes–large-, intermediate-, and small-scale fields–as in the investigation of the temporal correlation in Figure~\ref{fig2}(b). The decomposed fields predicted for the lead time of $T_L$ by the cGAN and CNN, for example, were compared with those of the target field, as shown in Figure~\ref{fig9}. As shown in Figure~\ref{fig2}(b), the large-scale field $(k\leq 4)$ persists longer than the total fields with an integral time scale 1.4 times larger than that of the total field, whereas the intermediate-scale field $(4<k\leq 20)$ and the small-scale field $(20<k)$ decorrelate more quickly than the total field with integral time scales one-fourth and one-twelfth of that of the total field, respectively. Given that the intermediate- and small-scale fields of the target field are almost completely decorrelated from the initial field at a lead time of $T_L$ as shown in Figure~\ref{fig2}(b), the predictions of those fields by the cGAN shown in Figures~\ref{fig9}(c) and \ref{fig9}(d) are excellent. The cGAN generator generates a small-scale field that not only mimics the statistical characteristics of the target turbulence but is also consistent with the large-scale motion of turbulence through adversarial learning. A comparison of the decomposed fields predicted by the cGAN and CNN for other lead times is presented in Appendix~\secref{appA}. For small lead times, such as $0.25T_L$ and $0.5T_L$, the cGAN predicted the small-scale fields accurately, whereas those produced by the CNN contained non-negligible errors. For $2T_L$, it is difficult to predict using both models even with the large-scale field. On the other hand, the CC of decomposed fields between the target and predictions, provided in Table~\ref{table4} of Appendix~\secref{appA}, did not demonstrate the superiority of the cGAN over the CNN. This is attributed to the fact that pointwise errors such as the MSE or CC are predominantly determined by the behaviour of large-scale motions. The pointwise MSE is the only error used in the loss function of the CNN, whereas the discriminator loss based on the latent variable is used in the loss function of the cGAN in addition to the MSE. This indicates that the latent variable plays an important role in predicting turbulent fields with multiscale characteristics.

\begin{figure}
	\centerline{\includegraphics[width=0.4\columnwidth]{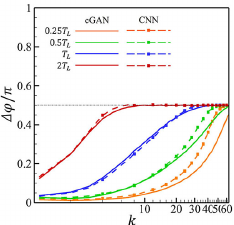}}
	\captionsetup{labelsep=period}
	\caption{Prediction results of the phase error of each model depending on $T$.}
	\label{fig10}
\end{figure}

In the prediction of the enstrophy spectrum, the cGAN showed excellent performance, compared to the CNN, by accurately reproducing the decaying behaviour of the spectrum in the small-scale range, as shown in Figure~\ref{fig8}. The average phase error between the target and predictions by the cGAN and CNN is shown in Figure~\ref{fig10}. For all lead times, the phase error of the small-scale motion approaches $\pi/2$, which is the value for a random distribution. For short lead times $0.25T_L$ and $0.5T_L$, the cGAN clearly suppressed the phase error in the small-scale range compared with the CNN, whereas for $T_L$ and $2T_L$, both the cGAN and CNN poorly predicted the phase of the intermediate- and small-scale motions, even though the cGAN outperformed the CNN in predicting the spectrum.

The performance of PredictionNet in the prediction of velocity fields is presented in Appendix~\secref{appB}, where several statistics, including the PDF of velocity, two-point correlation, and PDF of the velocity gradient are accurately predicted for all lead times. The superiority of the cGAN over the CNN was also confirmed.

\subsection{Role of the discriminator in turbulence prediction}\label{sub:pred2}
In the previous section, we demonstrated that adversarial learning using a discriminator network effectively captured the small-scale behaviour of turbulence, even when the small-scale field at a large lead time was hardly correlated with that of the initial field. In this section, we investigate the role of the discriminator in predicting turbulence through the behaviour of the latent variable, which is the output of the discriminator. Although several attempts have been made to disclose the manner in which adversarial training affects the performance of the generator and the meaning of the discriminator output~\citep{Creswell2018,Yuan2019,Goodfellow2020}, there have been no attempts to reveal the implicit role of the latent variable in recovering the statistical nature of turbulence in the process of a prediction. 

\begin{figure}
	\centerline{\includegraphics[width=1.0\columnwidth]{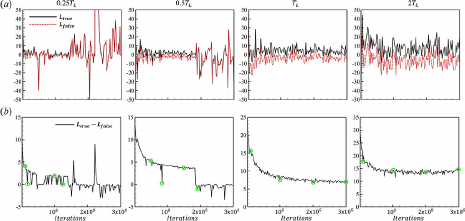}}
	\captionsetup{width=\columnwidth,justification=justified,labelsep=period}
	\caption{Evolution over training iteration of (a) $L_{true}$ and $L_{false}$, and (b) their difference evaluated every 2,000 iterations.}
	\label{fig11}
\end{figure}

In the current cGAN framework, the discriminator network is trained to maximise the difference between the expected value of the latent variable of the target field $D(Y,X)$ and that of the prediction $D(Y^*,X)$, whereas the generator network is trained to maximise $D(Y^*,X)$ and to minimise the pointwise MSE between $Y$ and $Y^*$. Therefore, through the adversarial learning process, the discriminator is optimised to distinguish $Y^*$ from $Y$, whereas the generator evolves to produce $Y^*$ by reflecting on the optimised latent variable of the discriminator and minimising the MSE between $Y$ and $Y^*$. The input field $X$ is used as a condition in the construction of the optimal latent variable in the discriminator network. The behaviour of the expected value of the latent variable of the target and generated fields ($L_{true}$ and $L_{false}$) and their difference ($L_{true}-L_{false}$) as learning proceeds for all lead times are illustrated in Figure~\ref{fig11}. Because of the regularisation of the latent variable by $L_{drift}(=L_{true}^2)$ in the loss function of the discriminator, both $L_{true}$ and $L_{false}$ remain around zero, although they sometimes oscillate significantly. As the learning proceeds, $L_{true}-L_{false}$ quickly decays initially owing to the suppression of the MSE and continues to decrease to zero for $0.25T_L$ and $0.5T_L$ but to a finite value for $T_L$ and $2T_L$. The intermittent transition of $L_{true}-L_{false}$ between a finite value and zero for $0.25T_L$ and $0.5T_L$ clearly suggests that the generator and discriminator operate competitively in an adversarial manner. As the generator improves, the difference monotonically decreases and occasionally exhibits sudden suppression. When sudden suppression occurs, the discriminator evolves to distinguish $Y^*$ from $Y$ by finding a better criterion latent variable, resulting in a sudden increase in the difference. Ultimately, when the generator can produce $Y^*$ that is indistinguishable from $Y$ by any criterion proposed by the discriminator, the difference converges to zero, as shown in the cases of $0.25T_L$ and $0.5T_L$. However, for $T_L$ and $2T_L$, such an event never occurs; $L_{true}-L_{false}$ monotonically decreases and tends toward a finite value, implying that the discriminator wins and can distinguish $Y^*$ from $Y$. Although the generator cannot produce $Y^*$ to beat the discriminator, $Y^*$ retains the best possible prediction.

\begin{figure}
	\centerline{\includegraphics[width=1.0\columnwidth]{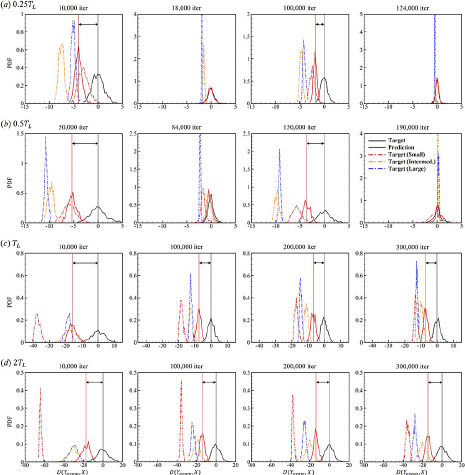}}
	\captionsetup{width=\columnwidth,justification=justified,labelsep=period}
	\caption{Distributions of discriminator output (latent variable) for various fields at several iterations. (a) $T=0.25T_L$, (b) $T=0.5T_L$, (c) $T=T_L$, and (d) $T=2T_L$. All distributions are shifted by the target mean to fix the target distribution at the zero mean. The vertical black solid line indicates the target mean (zero) and the red line is the prediction mean. When the mean difference between the target and prediction is smaller than 0.5, the vertical lines are omitted for visibility.}
	\label{fig12}
\end{figure}

To understand the mechanism in more detail that the discriminator uses to distinguish a generated field from a target field, the behaviour of the distribution of the latent variable during the learning process was investigated, because the latent variable plays a key role in the discriminator network. The distribution of the discriminator output of the generated field $D(Y^*,X)$ at four iterations during the learning process is marked with a green circle in Figure \ref{fig11}(b) for all lead times and is compared with that of the target field $D(Y,X)$ in Figure~\ref{fig12}.  The distributions of latent variable of the scale-decomposed target field $D(Y^S,X), D(Y^I,X)$, and $D(Y^L,X)$ are also compared for analysis, where $Y^S, Y^I$ and $Y^L$ are scale-decomposed from $Y$ in the same manner, as shown in Figure~\ref{fig2}.  Here, an additional 500 fields, in addition to the 50 test fields, were used to extract a smooth distribution of the latent variables of the target field and scale-decomposed target fields. For easy comparison, the mean value of the latent variable for the target field was shifted to zero. 

For all lead times, as learning proceeded, the distributions of $D(Y,X)$ and $D(Y^*,X)$ became narrower, with the mean values of $D(Y,X)$ and $D(Y^*,X)$ becoming closer to each other (the gap between the two vertical lines decreases). This indicates that, as the generator improves in producing $Y^*$ closer to $Y$, the discriminator becomes more efficient in distinguishing $Y^*$ from $Y$ because only the mean values are compared in the discrimination process and the narrower distributions yield the sharper mean values. Even when the mean values of $D(Y,X)$ and $D(Y^*,X)$ are almost the same (18,000 and 124,000 iterations for $0.25T_L$ and 84,000 and 190,000 iterations for $0.5T_L$, Figure~\ref{fig12}(a) and \ref{fig12}(b)), a more optimal discriminator at later iterations yields a narrower distribution. The collapse of the distributions of $D(Y,X)$ and $D(Y^*,X)$ in the second column for $0.25T_L$ and $0.5T_L$ occurs when the discriminator falls into a local optimum during training, implying that the predicted and target fields are indistinguishable by the discrimination criterion right at that iteration. However, as the criterion is jittered in the next iterations, the two fields become distinct again as shown in the third column. Eventually, when the two fields becom indistinguishable by any criterion after sufficient iterations, almost perfect collapse of very narrow distributions is achieved as shown in the fourth column for $0.25T_L$ (the collapse shown in the fourth column for $0.5T_L$ occurs in another local optimum). For $T_L$ and $2T_L$, for which perfect training of the discriminator was not achieved, however, distributions of $D(Y,X)$ and $D(Y^*,X)$ hardly change with iteration although the mean values are getting closer to each other very slowly.

In the comparison with the scale-decomposed target field, we observe that for $0.25T_L$ and $0.5T_L$, the distribution of the latent variable of the small-scale target field among all the scale-decomposed target fields is closest to that of the target field and generated field, whereas that of the intermediate-scale target field is closest to that of the target and generated fields for $T_L$ and $2T_L$. This suggests that a small-scale (or intermediate) field in the target field plays the most critical role in discriminating the generated field from the target field for $0.25T_L$ and $0.5T_L$ (or $T_L$ and $2T_L$). If the generator successfully produces $Y^*$ similar to $Y$ for $0.25T_L$ and $0.5T_L$, and the predictions for $T_L$ and $2T_L$ are incomplete, the small-scale fields for $T_L$ and $2T_L$ are useless for discriminating the generated field from the target field, and the intermediate-scale field is partially useful. Considering the possible distribution of the functional form of the latent variable may provide a clue to this conjecture. The latent variable, which is the output of the discriminator network shown in Figure~\ref{fig5}(b), and a function of the input fields $Y$ (or $Y^*$) and $X$, can be written as
\begin{equation}
D(Y,X) = \sum_{k_x, k_y} \widehat{D}^Y(k_x \Delta x, k_y \Delta y; w_d, \epsilon_L) \widehat{Y}(k_x,k_y) 
              +\sum_{k_x, k_y} \widehat{D}^X(k_x \Delta x, k_y \Delta y; w_d, \epsilon_L) \widehat{X}(k_x,k_y)
\end{equation}       
where $\widehat{Y}$ and $\widehat{X}$ are the Fourier coefficients of $Y$ and $X$, respectively. $\Delta x, \Delta y, w_d$, and $\epsilon_L$ are the grid sizes in each direction, weights of the discriminator network, and slope of the LReLU function for the negative input, respectively. This linear relationship between the input and output is a consequence of the linear structure of the discriminator network: the convolution, average pooling, and full connection operations are linear, and the leaky ReLU function is a piecewise linear function such that either one or $\epsilon_L$ is multiplied by the function argument, depending on its sign. Although $\widehat{D}^Y$ is an undetermined function, a possible shape can be conjectured. For $0.25T_L$ and $0.5T_L$, $\widehat{D}^Y$ of the optimised discriminator network has more weight in the small-scale range than in other scale ranges such that the latent variable is more sensitive to the small-scale field; thus, it better discriminates the generated field from the target field. Similarly, for $T_L$ and $2T_L$, $\widehat{D}^Y$ has more weight in the intermediate-scale range than in the other ranges, and discrimination is limited to the intermediate-scale field because the small-scale field is fully decorrelated; thus, it is no longer useful for discrimination. Although the manner in which $\widehat{D}^X$ conditionally influences learning is unclear, the scale-dependent correlation of the target field with the initial input field appears to be captured by $\widehat{D}^X$. This scale-selective feature of the discriminator appears to be the key element behind the successful prediction of the small-scale characteristics of turbulence. Here, the generator implicitly learns system characteristics embedded in data to deceive the discriminator with such features; thus, it's extremely challenging to provide an exact mechanism for how prediction performance is comprehensively enhanced, particularly in terms of physical and statistical attributes through adversarial training. However, we clearly showed that such a successful prediction of turbulence, even down to small scales, is possible through adversarial learning and not by the use of a CNN, which enforces suppression of the pointwise MSE only.

One last question that may arise in this regard is whether introducing physical constraints explicitly into the model, rather than using implicit competition with an additional network, could also be effective. To explore this, we incorporated the enstrophy spectrum as an explicit loss term into the objective function of the CNN model to address the performance issues associated with small-scale features. As shown in the relevant results presented in Appendix~\secref{appC}, adding the enstrophy loss alone did not lead to better performance although the enstrophy spectrum was better predicted. This confirms that the adoption of an implicit loss based on the latent variable in cGAN is effective in reflecting the statistical nature of turbulence, particularly the small-scale characteristics.

\subsection{PredictionNet – single vs recursive prediction}\label{sub:pred3}
In the previous section, we observed that the prediction performance for large lead times deteriorated naturally because of the poor correlation between the target field and the initial input field. An improved prediction for large lead times may be possible if recursive applications of the prediction model for short lead times are performed. Conversely, recursive applications may result in the accumulation of prediction errors. Therefore, there is an optimal model that can produce the best predictions. In this regard, since the performance of CNN models itself falls short of PredictionNet across all lead times, a more significant error accumulation is expected to arise. Thus, we would focus on analyzing the recursive prediction results using PredictionNet. Nevertheless, we also presented the results of CNN models, not only to highlight the improvements through recursive prediction for large lead-time prediction but also to compare how much further enhancement is achieved when utilizing PredictionNet. A similar concept was tested for weather prediction~\citep{Daoud2003}.

\begin{figure}
	\centerline{\includegraphics[width=0.9\columnwidth]{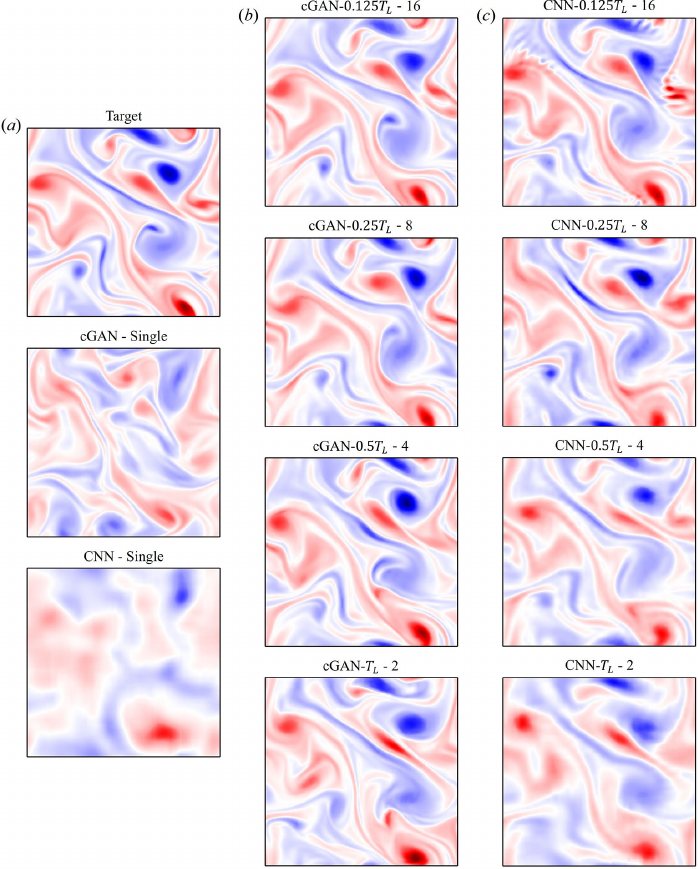}}
	\captionsetup{width=\columnwidth,labelsep=period}
	\caption{Visualised prediction results for $T=2T_L$ of (a) target DNS field and single predictions using cGAN and CNN, (b) cGAN recursive predictions, and (c) CNN recursive predictions.}
	\label{fig13}
\end{figure}

\begin{table}\centering
	\begin{tabular}{c c c c c c| c c c c| c c}
		\multirow{2}{*}{\makecell[c]{lead time\\$(T)$}} & \multirow{2}{*}{\makecell[c]{recursive\\ model}} & \multicolumn{2}{c}{CC} & \multicolumn{2}{c|}{MSE} & \multicolumn{2}{c}{RMS} & \multicolumn{2}{c|}{$\hat{\mu}_4$} & \multicolumn{2}{c}{$\varepsilon' (\times 10^{-2})$} \\ \cmidrule{3-12}\addlinespace[3pt]
		& & cGAN & CNN & cGAN & CNN & cGAN & CNN & cGAN & CNN & cGAN & CNN \\
		\hline
		\multirow{6}{*}{$2T_L$} & Target & \multicolumn{2}{c}{-} & \multicolumn{2}{c|}{-} & \multicolumn{2}{c}{0.788} & \multicolumn{2}{c|}{3.65} & \multicolumn{2}{c}{1.28} \\[1pt]
		& $0.125T_L$-16 & 0.625 & 0.757 & 0.341 & 0.253 & 0.736 & 0.764 & 3.51 & 3.28 & 1.12 & 1.20 \\[1pt]
		& $0.25T_L$-8 & 0.672  & 0.558 & 0.300 & 0.419 & 0.746 & 0.722 & {\bf 3.66} & 3.59 & 1.15 & 1.07 \\[1pt]
		& $\boldsymbol{ 0.5T_L}${\bf-4} & {\bf 0.739} & 0.705 & {\bf 0.250} & 0.259 & {\bf 0.776} & 0.625 & 3.75 & 3.52 & {\bf 1.24} & 0.80 \\[1pt]
		& $T_L$-2 & 0.708 & 0.766 & 0.281 & 0.209 & 0.758 & 0.656 & 3.61 & 3.53 & 1.18 & 0.88 \\[1pt]
		& Single & 0.453 & 0.523 & 0.506 & 0.386 & 0.698 & 0.580 & 3.57 & 4.06 & 1.00 & 0.70 \\
		\hline
		\multirow{5}{*}{$T_L$} & Target & \multicolumn{2}{c}{-} & \multicolumn{2}{c|}{-} & \multicolumn{2}{c}{0.888} & \multicolumn{2}{c|}{3.27} & \multicolumn{2}{c}{1.62} \\[1pt]
		& $0.125T_L$-8 & 0.891 & 0.941 & 0.126 & 0.0708 & 0.854 & 0.828 & 3.20 & 3.12 & 1.50 & 1.41 \\[1pt]
		& $0.25T_L$-4 & 0.916 & 0.866 & 0.0971 & 0.157 & 0.860 & 0.819 & 3.30 & 3.34 &1.52 & 1.38 \\[1pt]
		& $\boldsymbol{0.5T_L}${\bf-2} & {\bf 0.924} & 0.919 & {\bf 0.0907} & 0.0936 & {\bf 0.875} & 0.751 & 3.32 & 3.30 & {\bf 1.57} & 1.16 \\[1pt]
		& Single & 0.855 & 0.887 & 0.172 & 0.130 & 0.860 & 0.795 & {\bf 3.26} & 3.41 & 1.52 & 1.30 \\
		\hline
		\multirow{4}{*}{$0.5T_L$} & Target & \multicolumn{2}{c}{-} & \multicolumn{2}{c|}{-} & \multicolumn{2}{c}{0.944} & \multicolumn{2}{c|}{3.08} & \multicolumn{2}{c}{1.83} \\[1pt]
		& $0.125T_L$-4 & 0.979 & 0.983 & 0.0277 & 0.0228 & 0.924 & 0.887 & 3.06 & 3.18 & 1.76 & 1.62 \\[1pt]
		& $\boldsymbol{0.25T_L}${\bf-2} & {\bf 0.983} & 0.971 & {\bf 0.0227} & 0.0383 & 0.928 & 0.882 & {\bf 3.11} & 3.29 & 1.77 & 1.60 \\[1pt]
		& Single & 0.971 & 0.970 & 0.0390 & 0.0404 & {\bf 0.936} & 0.892 & 3.11 & 3.15 & {\bf 1.80} & 1.64 \\
		\hline
		\multirow{3}{*}{$0.25T_L$} & Target & \multicolumn{2}{c}{-} & \multicolumn{2}{c|}{-} & \multicolumn{2}{c}{0.973} & \multicolumn{2}{c|}{3.00} & \multicolumn{2}{c}{1.94} \\[1pt]
		& $\boldsymbol{0.125T_L}${\bf-2} & {\bf 0.995} & 0.993 & {\bf 7.41e-3} & 9.51e-3 & 0.962 & 0.921 & {\bf 2.99} & 3.21 & 1.90 & 1.75 \\[1pt]
		& Single & 0.994 & 0.990 & 8.58e-3& 0.0144 & {\bf 0.964} & 0.960 & 3.02 & 3.02 & {\bf 1.91} & 1.90\\
		\hline
		\multirow{2}{*}{$0.125T_L$} & Target & \multicolumn{2}{c}{-} & \multicolumn{2}{c|}{-} & \multicolumn{2}{c}{0.987} & \multicolumn{2}{c|}{2.96} & \multicolumn{2}{c}{2.00} \\[1pt]
		& Single & 0.998 & 0.997 & 2.74e-3 & 4.75e-3 & 0.981 & 0.981 & 2.96 & 2.96 & 1.98 & 1.98 \\
		\hline
	\end{tabular}
	\captionsetup{labelsep=period}
	\caption{Statistical results of recursive predictions for various $T$'s including the single prediction at the last entries. Best prediction for each lead time by cGAN is bold-faced.}
	\label{table3}
\end{table}

\begin{figure}
	\centerline{\includegraphics[width=0.85\columnwidth]{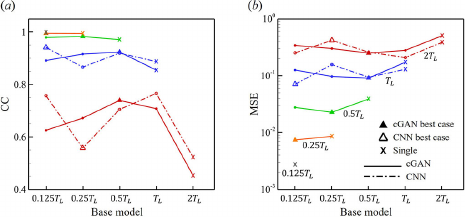}}
	\captionsetup{labelsep=period}
	\caption{Plots of the CC and MSE of the recursive predictions in terms of the lead time of the base recursive model. Only $T_L$ and $2T_L$ cases are included for CNN for visibility.}
	\label{fig14}
\end{figure}

\begin{figure}
	\centerline{\includegraphics[width=0.85\columnwidth]{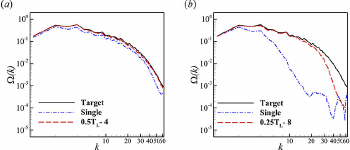}}
	\captionsetup{labelsep=period}
	\caption{Enstrophy spectra of the best case of recursive prediction compared with the target and single prediction for $T=2T_L$ using (a) cGAN and (b) CNN.}
	\label{fig15}
\end{figure}

For example, for the prediction of a large lead time $2T_L$, four base models trained to predict the lead times $0.125T_L, 0.25T_L, 0.5T_L$, and $T_L$ were recursively applied 16, 8, 4, and 2 times, respectively, and compared against the single prediction as shown in Figure~\ref{fig13}. All recursive applications produced better predictions than the corresponding single prediction, and among them, four recursive applications of the cGAN model for $0.5T_L$ yielded the best results. Eight recursive applications of CNN model for $0.25T_L$, however, produces the best prediction. For all lead times, the best prediction was sought; the performances of all recursive predictions by cGAN and CNN are compared in Table~\ref{table3} in terms of performance indices, such as the CC and MSE and statistical quantities, including RMS, $\hat{\mu}_4$, and dissipation rate. Here, we observed that the performance difference depending on the input time-point varies significantly depending on the base model; thus, the time-averaged values for CC and MSE from $t_0$ to $t_{100}$ as input are presented to ensure fair comparison, unlike \secref{sub:pred1}. The CC and MSE values are plotted in Figure~\ref{fig14}. As expected, there exists an optimum base model producing the best performance for each lead time. Recursive prediction was more effective for large lead times ($T_L$ and $2T_L$) than for short lead times ($0.25T_L$ and $0.5T_L$) for cGAN model. For $2T_L$, the CC improved from 0.4530 to 0.7394 and MSE decreased from 0.506 to 0.250 through the best recursive applications of cGAN model.
The predicted statistics exhibited a consistent improvement. For $T_L$, the recursive applications show similar improvements. However, for $0.25T_L$ and $0.5T_L$, even though the recursive applications produced slightly better performance in terms of the CC and MSE, the statistics predicted by the single prediction are more accurate than those of the recursive prediction. However, recursive applications of CNN model do not show a monotonic behavior in CC or MSE; for $T_L$, eight applications of CNN model trained for $0.125T_L$ yield the best prediction, whereas two applications of CNN model for $T_L$ produces the best prediction for $2T_L$. Finally, the prediction of the enstrophy spectrum by recursive prediction also yielded an improvement over the single prediction, as shown in Figure~\ref{fig15}. Particularly, it is noticeable that eight recursive applications of CNN model trained for $0.25T_L$ yielded relatively better spectrum than the single prediction. However, the performance of prediction was not good enough as shown in Figure~\ref{fig13}(c).

\subsection{ControlNet - maximisation of the propagation of the control effect}\label{sub:control1}
In this section, we present an example of flow control that uses the high-accuracy prediction model developed in the previous section as a surrogate model. By taking advantage of the prediction capability of the model, we can determine the optimum control input that can yield the maximum modification of the flow in the specified direction with a fixed control input strength. In particular, we trained ControlNet to find a disturbance field that produces the maximum modification in the vorticity field over a finite time period with the constraint of a fixed RMS of the disturbance field. Similar attempts to control 2D decaying turbulence were made by \citet{Jimenez2018} and \citet{Yeh2021}. \citet{Jimenez2018} modified a part of a vorticity field to identify dynamically significant sub-volumes in 2D decaying turbulence. The influence on the future evolution of the flow was defined as significance using the L2 norm of the velocity field. Then, the significance of the properly-labelled regions was compared by turning the vorticity of specific regions on and off. They showed that vortices or vortex filaments are dynamically significant structures, and interstitial strain-dominated regions are less significant. In contrast, \citet{Yeh2021} developed a method called network-broadcast analysis based on network theory by applying Katz centrality \citep{Katz1953} to identify key dynamical paths along which perturbations amplify over time in 2D decaying turbulence. Two networks (composed of nodes and edges, not neural networks), the Biot–Savart network (BS) and Navier–Stokes network (NS), with different adjacency matrices were investigated. In the former case, vortex structures similar to those in \citet{Jimenez2018} and in the latter case, opposite-sign vortex pairs (vortex dipoles) were the most effective structures for perturbation propagation. However, both studies confined the control disturbance field either by localising the modification for control by dividing the domain into cell units \citep{Jimenez2018}, or by considering the perturbation using the leading singular vector of an adjacency matrix calculated from a pulse in a predefined shape \citep{Yeh2021}. However, the disturbance field considered in ControlNet is free of shape, and the only constraint is the strength of the disturbance field in terms of the fixed RMS of the disturbance field. 

\begin{figure}
	\centerline{\includegraphics[width=0.45\columnwidth]{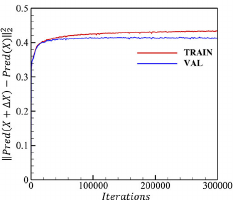}}
	\captionsetup{labelsep=period}
	\caption{Convergence of the data loss of ControlNet for $0.5T_L$. $\Delta X_{rms}$=0.1$X_{rms}$.}
	\label{fig16}
\end{figure}

As a surrogate model for the time evolution of the disturbance-added initial field, PredictionNet trained for a lead time of $0.5T_L$ showing excellent prediction, was selected for the control test. ControlNet is trained to generate an optimum disturbance field $\Delta X$ that maximises the difference between the disturbance-added prediction $\tilde{Y} (=Pred(X+\Delta X))$ and the original (undisturbed) prediction $Y^*$. If the strength of $\Delta X$ is too large, causing $X+\Delta X$ to deviate from the range of dynamics of the pre-trained PredictionNet, the surrogate model will not function properly, resulting in $\tilde{Y}$ being different from the ground-truth solution $\mathscr{N}(X+\Delta X)$. Here, $\mathscr{N}()$ is the result of the Navier--Stokes simulation, with the input as the initial condition. $\tilde{Y}$ using disturbances with various strengths of the RMS of the input field ($\Delta X_{rms}$=0.1, 0.5, 1, 5, 10, 20\% of $X_{rms}$) were compared with the corresponding $\mathscr{N}(X+\Delta X)$. We confirmed that PredictionNet functions properly within the tested range (Appendix~\secref{appD}). Therefore, the effect of the disturbance strength was not considered; thus, we only present the 10\% case. Coefficient $\theta$ related to the smoothing effect was fixed at 0.5 for $\Delta X_{rms}=0.1X_{rms}$ through parameter calibration. Figure~\ref{fig16} shows the data loss trend of ControlNet for $\Delta X_{rms}=0.1X_{rms}$, which is maximised as the training progresses, confirming successful training.

\begin{figure}
	\centerline{\includegraphics[width=1.0\columnwidth]{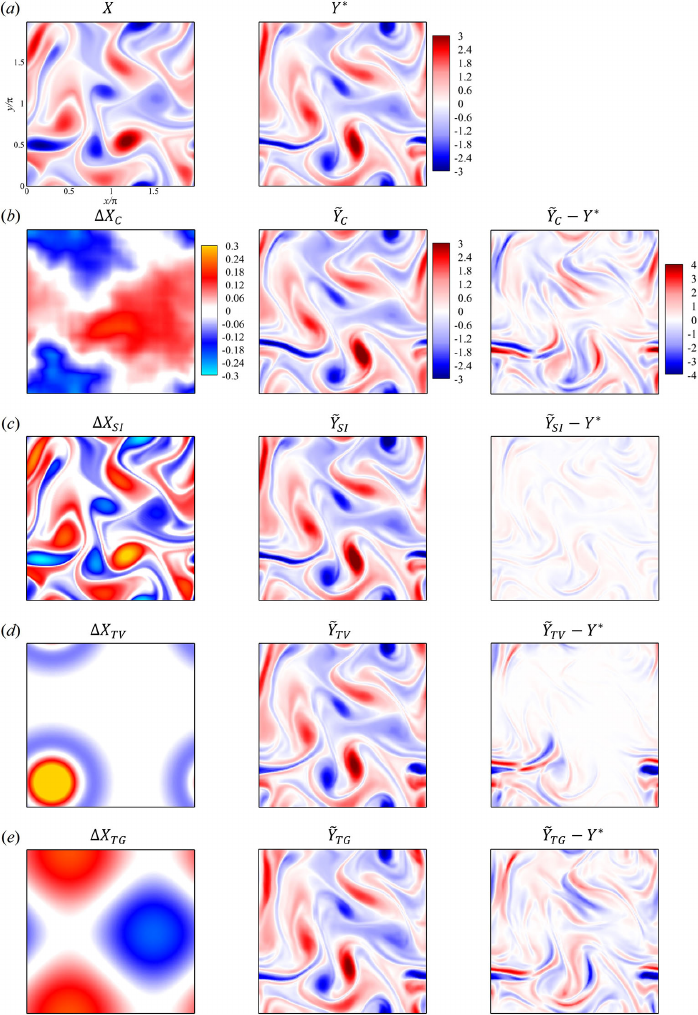}}
	\captionsetup{width=\columnwidth,justification=justified,labelsep=period}
	\caption{Visualised example of disturbance fields. (a) Input and the undisturbed prediction, (b) the optimum disturbance ($Control(X)$), disturbance-added prediction ($Pred(X+\Delta X_C)$), and the difference. Comparison cases of (c) $\Delta X_{SI}$, (d) $\Delta X_{TV}$, and (e) $\Delta X_{TG}$.}
	\label{fig17}
\end{figure}

Figure~\ref{fig17} presents a visualisation of the effect of the control input on one of the test data. Figure~\ref{fig17}(a) shows the input vorticity field and corresponding prediction at $T=0.5T_L$. Figure~\ref{fig17}(b) shows the disturbance field generated by ControlNet  $\Delta X_C$, disturbance-added prediction, and difference between the undisturbed and disturbance-added predictions, demonstrating that the control effect is effectively propagated and amplified at $T=0.5T_L$. The MSE between $\tilde{Y}_{C}$ and $Y^*$ is 0.757, indicating a substantial change. A prominent feature of $\Delta X_{C}$ is its large-scale structure.
To verify whether the propagation of the control effect of $\tilde{Y}_C$ was maximised under the given conditions, we considered three additional disturbance fields. Inspired by the results of \citet{Jimenez2018} and the claim of \citet{Yeh2021} regarding their BS network, in which vortices have the greatest effect on the future evolution of the base flow, we considered a 10\% scaled input disturbance $\Delta X_{SI} (=0.1 X)$. Another disturbance field due to a single Taylor vortex $\Delta X_{TV}$ extracted from the NS network of \citet{Yeh2021}, which is best for amplifying the control effect when applied to a vortex dipole rather than the main vortex structures, is considered in the following form: 
\begin{equation}\label{eq4.2}
	\Delta X_{TV}(x,y)=\epsilon\delta(x_c,y_c)= \epsilon(2/r_{\delta}-r^2/r^3_{\delta})\mathrm{exp}[-r^2/(2r^2_{\delta})],
\end{equation}
where $r=\sqrt{(x-x_c)^2+(y-y_c)^2}$ with vortex center $(x_c,y_c)$. $\epsilon$ represents the amplitude of the Taylor vortex, and it is adjusted to maintain the RMS of the disturbance at 0.1$X_{rms}$. $r_{\delta}$ was set to $r_{\delta}=\pi/k_{max}$, where $k_{max}=4$ denotes the wavenumber with the maximum value in the enstrophy spectra.
Finally, the disturbance field in the form of a Taylor--Green vortex  $\Delta X_{TG}$ approximating the disturbance field of ControlNet is considered as follows. 
\begin{equation}\label{eq4.3}
	\Delta X_{TG}(x,y)=\epsilon \sin \left(\frac{1}{\sqrt{2}}(x-x_c-y+y_c)\right) \sin \left(\frac{1}{\sqrt{2}}(x-x_c+y-y_c)\right),
\end{equation}
indicating the largest vortex in the $[0,2\pi)^2$-domain.
For $\Delta X_{TV}$ and $\Delta X_{TG}$, the optimum location of $(x_c,y_c)$ that yielded the greatest propagation was determined through tests using PredictionNet.

 \begin{figure}
 	\centerline{\includegraphics[width=0.9\columnwidth]{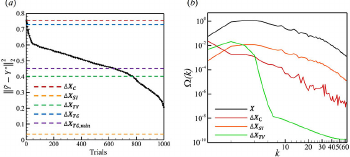}}
 	\captionsetup{width=\columnwidth,justification=justified,labelsep=period}
 	\caption{(a) Comparison of MSEs between various $\tilde{Y}$ and $Y^*$ using different disturbance fields. Black dots denote the sorted MSE of phase-shifted $\Delta X_C$. The first node of black dots (red dashed line) shows the result of the optimum $\Delta X_{C}$. (b) Enstrophy spectra of the input $X$, $\Delta X_{C}$, $\Delta X_{SI}$, and $\Delta X_{TV}$.}
 	\label{fig18}
 \end{figure}
 
The disturbance-added predictions $\tilde{Y}_{SI}$ and $\tilde{Y}_{TV}$ and their differences from the undisturbed prediction $Y^*$ using $\Delta X_{SI}$ and $\Delta X_{TV}$ located at the optimum position as disturbances, respectively, are shown in Figures~\ref{fig17}(c) and \ref{fig17}(d). The difference fields show that these disturbances do not significantly change the field. MSE values for $\tilde{Y}_{SI}$ and $\tilde{Y}_{TV}$ against the undisturbed prediction are 0.039 and 0.402, respectively, which are much smaller than the corresponding value of 0.757 for $\tilde{Y}_{C}$, as shown in Figure~\ref{fig18}(a). From this comparison, it can be conjectured that because the goal of control is to maximise the pointwise MSE of the vorticity against the undisturbed prediction, the large-scale disturbance of the vorticity is more effective in displacing and deforming the vorticity field. The enstrophy spectra of the disturbance fields shown in Figure~\ref{fig18}(b) confirm that $\Delta X_C$ has a peak value at $k=1$, representing the largest permissible scale in the given domain, whereas $\Delta X_{TV}$ and $\Delta X_{SI}$ have peak values at $k=2$ and $k=4$, respectively, under the constraint that the RMS value of all disturbances is $0.1X_{rms}$. This observation leads to the consideration of the largest Taylor--Green vortex-type disturbance $\Delta X_{TG}$ in the domain given by Equation~(\ref{eq4.3}). The distribution of the optimum location of $\Delta X_{TG}$ shown in Figure~\ref{fig17}(e), yielding the greatest propagation coincides with that of $\Delta X_C$ shown in Figure~\ref{fig17}(b) (except for the sign owing to the symmetry in the propagation effect, as shown in the third panel of Figure~\ref{fig17}(b) and \ref{fig17}(e)). The MSE of $\tilde{Y}_{TG}$ against $Y^*$ was 0.726, which was slightly smaller than 0.757 of $\tilde{Y}_C$ (Figure~\ref{fig18}(a)). All these comparisons suggest that the largest-scale disturbance is the most effective in modifying the vorticity field through displacement or distortion of the given vorticity field. To verify whether the location of the largest-scale disturbance $\Delta X_C$ was indeed globally optimal, tests were conducted with a phase-shifted $\Delta X_C$. We consider two types of phase shifting for the wavenumber components $\widehat{\Delta X}_C \exp (i k_x l_x + i k_y l_y)$: one with randomly selected $l_x$ and $l_y$ uniformly applied to all wavenumbers and the other with randomly selected phases differently applied to each wavenumber. The test results confirm that the maximum MSE was obtained for $\Delta X_C$ (without a phase shift), with the minimum MSE found at 0.2 among the 1,000 trial disturbances considered with randomly selected phases, as shown in Figure~\ref{fig18}(a). $\Delta X_{TG,min}$ corresponds to one of $\Delta X_{TG}$ tests yielding the minimum modification with an MSE of approximately 0.45. The wide range of MSE for the largest-scale disturbance indicates that the location of the largest-scale disturbance is important for modifying the flow field.

\begin{figure}
	\centerline{\includegraphics[width=1.0\columnwidth]{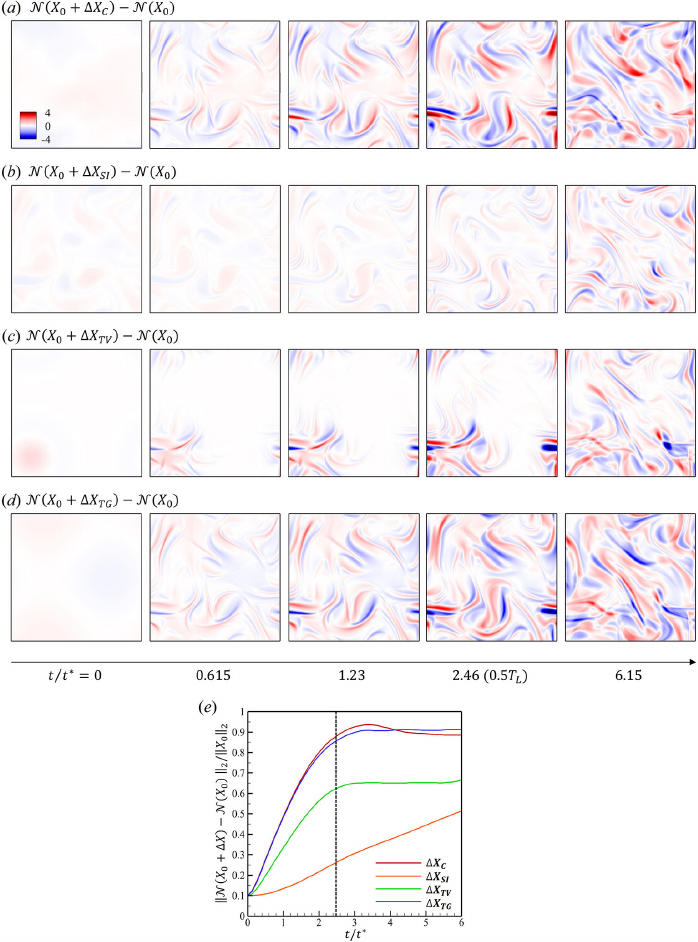}}
	\captionsetup{width=\columnwidth,justification=justified,labelsep=period}
	\caption{Simulated propagations of the control effect along with the time horizon presented by differences between the disturbance-added simulations and the original simulation. (a) $\Delta X_{C}$, (b) $\Delta X_{SI}$, (c) $\Delta X_{TV}$, and (d) $\Delta X_{TG}$. The RMSE result of the propagation of each disturbance is shown in (e) after being normalised by the input RMS.}
	\label{fig19}
\end{figure}

To confirm the optimality of $\Delta X_C$ in real flow, full Navier--Stokes simulations were performed further than $0.5T_L$ using the disturbance-added initial conditions for $\Delta X_{C}$, $\Delta X_{SI}$, $\Delta X_{TV}$, and $\Delta X_{TG}$. The visualisation results of the propagation over the time horizon are shown in Figure~\ref{fig19}, and the behaviour of the normalised RMSE over time is shown in Figure~\ref{fig19}(e). Therefore, the propagation by large-scale disturbances such as $\Delta X_{C}$ and $\Delta X_{TG}$ is much more effective than that by $\Delta X_{SI}$ and $\Delta X_{TV}$ even up to longer than $T_L$. Moreover, the surrogate model functions properly for $\Delta X_{rms}=0.1X_{rms}$ based on the comparison between the results at $0.5T_L$ in Figure~\ref{fig19} and those in the third column (difference fields) in Figure~\ref{fig17} for each disturbance. The RMSE of $\Delta X_C$ and $\Delta X_{TG}$ behave almost indistinguishably as shown in Figure~\ref{fig19}(e) for all test periods.

\begin{figure}
	\centerline{\includegraphics[width=0.85\columnwidth]{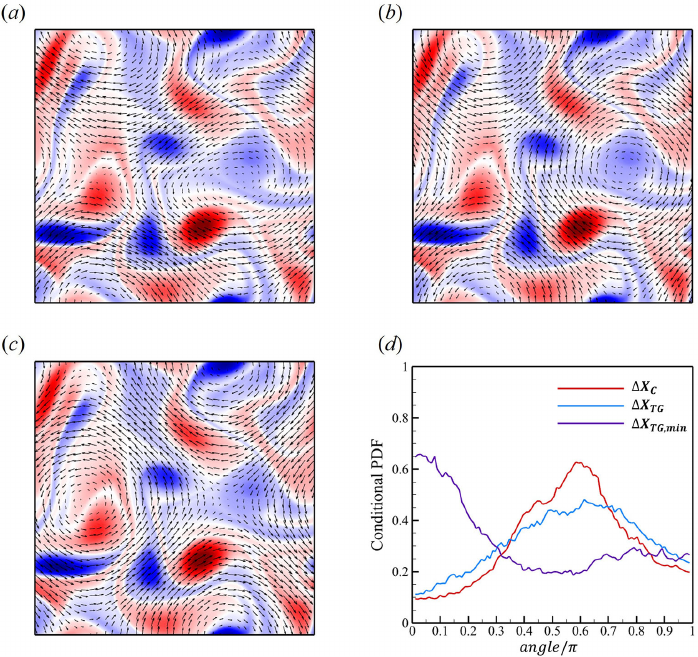}}
	\captionsetup{width=\columnwidth,justification=justified,labelsep=period}
	\caption{Distribution of the disturbance vector field with the input vorticity contours for (a) $\Delta X_{C}$, (b) $\Delta X_{TG}$, and (c) $\Delta X_{TG,min}$. (d) Conditional PDF of the angle between the velocity vectors of the input and disturbance.}
	\label{fig20}
\end{figure}

As shown in Figure~\ref{fig18}(a), the location of a large-scale disturbance causes a difference in the modification of the flow field. To investigate this in more detail, the vorticity contours of the flow field and the velocity vectors of the disturbances yielding the greatest change, $\Delta X_C$ and $\Delta X_{TG}$, and the disturbance producing the least change, $\Delta X_{TG,min}$ are plotted together in Figure~\ref{fig20}. The velocity vector distributions of $\Delta X_C$ (Figure~\ref{fig20}(a)) and $\Delta X_{TG}$ (Figure~\ref{fig20}(b)) are similar, whereas those of $\Delta X_{TG,min}$ (Figure~\ref{fig20}(c)) differ significantly. Careful observation shows that the maximum modification occurs when the velocity of the disturbances is applied in the direction normal to the elongation direction of the strong vortex region, whereas the flow field is minimally changed when the velocity of the disturbances is in the elongation direction of the strong vortex patch. The conditional PDF of the angle between the velocity vectors of the input field $X$ and disturbance fields is obtained under the condition that the velocities of the input and disturbance are greater than their own RMS values. Figure~\ref{fig20}(d) shows that the peak is found at approximately $0.6 \pi$ for $\Delta X_C$ and $\Delta X_{TG}$, and zero for $\Delta X_{TG,min}$, confirming this trend, given that the velocity vectors circumvent the vortex patches.

\section{Conclusion}\label{sec:conclusion}
In this study, the dynamics prediction of 2D DHIT was performed using a cGAN-based deep learning model. The developed prediction network, called PredictionNet, produced highly accurate predictions up to a lead time of half the Eulerian integral time scale. A quantitative comparison of the prediction performance with that of the baseline CNN model proved the superiority of the GAN-based model. In particular, the small-scale turbulence characteristics, which were not properly predicted by the CNN model, were captured well by the cGAN-based model. A detailed investigation of the behaviour of the latent variable, which is the output of the discriminator network, suggests that the discriminator network evolves through training to possess a scale-selection capability; thus, it can effectively distinguish the generated turbulence from the true turbulence. The minimisation of the pointwise mean-squared error loss tended to capture large-scale motions only, whereas competitive optimisation of the discriminator network using the latent variable led to the recovery of small-scale motions.

In addition, recursive predictions were tested using high-accuracy base models for short lead times to improve the predictive accuracy for long lead times. As shown in Figure~\ref{fig13}, four recursive applications of the prediction model trained for $0.5T_L$ yielded significantly better predictions than the single prediction for $2T_L$. However, more recursive applications of the prediction model for shorter lead times did not always guarantee improvement, indicating that an optimum recursive model exists for each lead-time prediction. 

Flow control was conducted as an example of application of the developed high-accuracy prediction model. Using the developed prediction model for $0.5T_L$ as a surrogate model, a control network was trained to provide an optimum disturbance field that maximised the modification at a given lead time. When the pointwise mean-squared difference between the disturbance-added prediction and undisturbed prediction at $0.5T_L$ was maximised, the optimum disturbance turned out to be of the largest scale fitting the domain. This maximum propagation of the control effect appeared to be achieved through the translation of elongated vortices in a direction orthogonal to the elongation direction. Although the optimum disturbances were found in a model-free manner (the only constraint is the condition for RMS), our models converged well to the optimum solution despite the high probability of falling into local optima because of the high degree of freedom. Although the control of 2-D turbulence using the distributed disturbances seems impractical, we provided an example of deep learning framework on which  using the well-trained prediction model, it is possible to find an optimal control input very efficiently that can change the flow as one wishes. It can easily extended to more practical applications.

Our investigation was restricted to 2D decaying homogeneous isotropic turbulence. Therefore, a natural extension would be toward more general turbulent flows such as inhomogeneous 2D flows or even 3D turbulence. Application to 3D homogeneous isotropic turbulence would be straightforward, even though there is a cost issue because training would require much more time than in 2D flows. However, it is worthwhile investigating whether the scale-selection capability of the discriminator network works in 3D isotropic turbulence. We have demonstrated that a generative model such as GAN is more effective in learning turbulence characteristics than a CNN. However, GAN is not the only generative model. Recently, a diffusion model \citep{Dickstein2015, Ho2020, Shu2023} was proposed as a generative model in which training is much more stable than GAN-based model. A comparative study of GAN and the diffusion models in the prediction of turbulence would be an interesting topic for future research.

	\vspace{0.3in}
	\noindent
	{\bf {Acknowledgments}}
	
	This work was supported by the National Research Foundation of Korea (NRF) grant funded by the Korean government (MSIP) (2022R1A2C2005538).
	
	\vspace{0.3in}
	\noindent
	{\bf Declaration of interest }. The authors declare no conflicts of interest.

\appendix
\section{Pseudo-spectral approximation of the vorticity-streamfunction formulation}\label{app0}
In two dimension, the governing equations for the vorticity-streamfunction formulation are
\begin{eqnarray}
\frac{\partial \omega}{\partial t} + \frac{\partial (u \omega)}{\partial x} + \frac{\partial (v \omega)}{\partial y} & = & \nu \left( \frac{\partial^2 \omega}{\partial x^2} + \frac{\partial^2 \omega}{\partial y^2} \right) \label{eqa1} \\
\frac{\partial^2 \psi}{\partial x^2} + \frac{\partial^2 \psi}{\partial y^2} & = & - \omega , \label{eqa2}
\end{eqnarray}
with
\begin{equation}
u = \frac{\partial \psi}{\partial y},~~ v = - \frac{\partial \psi}{\partial x} . \label{eqa3}
\end{equation}
In a biperiodic domain, the discrete Fourier representation of vorticity field is
\begin{equation}
\omega (x,y) = \sum_{\kappa_x} \sum_{\kappa_y} \widehat{\omega} (\kappa_x, \kappa_y, t) \exp (i \kappa_x x) \exp (i \kappa_y y)
\end{equation}
where $\widehat{\omega}$ is the Fourier coefficient of vorticity, and $\kappa_x$ and $\kappa_y$ are wavenumbers in $x-$ and $y-$directions, respectively. Pseudo-spectral approximation to Equation (\ref{eqa1})  yields
\begin{equation}
\frac{\partial \widehat{\omega}}{\partial t} + i \kappa_x \widehat{u \omega} + i \kappa_y \widehat{v \omega} = -\nu \kappa^2 \widehat{\omega} \label{eqa5}
\end{equation}
with
\begin{equation}
-\kappa^2 \widehat{\psi} = -\widehat{\omega}, ~~ \widehat{u} = i \kappa_y \widehat{\psi}, ~~\widehat{v}=- i \kappa_x \widehat{\psi} \label{eqa6}
\end{equation}
where $\widehat{\psi}, \widehat{u}$ and $\widehat{v}$ correspond to the Fourier coefficients of $\psi, u$ and $v$, respectively, and $\kappa^2 = \kappa_x^2+\kappa_y^2$. $\widehat{u\omega}$ and $\widehat{v \omega}$ are the Fourier coefficients of $u \omega$ and $v \omega$ which are pointwisely calculated in the physical space. From Equation (\ref{eqa6}),
\begin{equation}
\widehat{u}=\frac{i\kappa_y}{\kappa^2} \widehat{\omega}, ~~\widehat{v}=-\frac{i\kappa_x}{\kappa^2} \widehat{\omega},
\end{equation}
yielding with Equation (\ref{eqa5})
\begin{eqnarray}
\frac{\partial \widehat{u}}{\partial t} & = & \frac{i\kappa_y}{\kappa^2} \frac{\partial \widehat{\omega}}{\partial t} \\
  & = & \frac{i \kappa_y}{\kappa^2} \left( - i\kappa_x \widehat{u \omega} - i \kappa_y \widehat{v \omega} - \nu \kappa^2 \widehat{\omega} \right) \\
  & = & \widehat{v \omega} - i\kappa_x \left( -\frac{i \kappa_x}{\kappa^2} \widehat{v \omega} + \frac{i \kappa_y}{\kappa^2} \widehat{u \omega} \right) - \nu \kappa^2 \widehat{u} . \label{eqa10}
\end{eqnarray}
Similarly,
\begin{eqnarray}
\frac{\partial \widehat{v}}{\partial t} & = & - \frac{i\kappa_x}{\kappa^2} \frac{\partial \widehat{\omega}}{\partial t} \\
  & = & - \widehat{u \omega} - i\kappa_y \left( -\frac{i \kappa_x}{\kappa^2} \widehat{v \omega} + \frac{i \kappa_y}{\kappa^2} \widehat{u \omega} \right) - \nu \kappa^2 \widehat{v} . \label{eqa12}
\end{eqnarray}
Noticing that the quantity in the bracket in Equations (\ref{eqa10}, \ref{eqa12}) is identical, taking inverse Fourier transform of  both equations leads to
\begin{eqnarray}
\frac{\partial u}{\partial t} - v \omega & = & - \frac{\partial P}{\partial x} + \nu \left( \frac{\partial^2 u}{\partial x^2} +\frac{\partial^2 u }{\partial y^2} \right) , \label{eqa13} \\
\frac{\partial v}{\partial t} + u \omega & = & - \frac{\partial P}{\partial y} + \nu \left( \frac{\partial^2 v}{\partial x^2} +\frac{\partial^2 v }{\partial y^2} \right) , \label{eqa14}
\end{eqnarray}
with
\begin{equation} 
\widehat{P} = -\frac{i \kappa_x}{\kappa^2} \widehat{v \omega} + \frac{i \kappa_y}{\kappa^2} \widehat{u \omega} ,
\end{equation}
guaranteeing that the velocity field satisfies the divergence-free condition.
Equations (\ref{eqa13}, \ref{eqa14}) are the rotational form of the two-dimensional Navier--Stokes equation. This proves that the pseudo-spectral approximation to the vorticity-streamfunction formulation is equivalent to pseudo-spectral approximation to the rotational form of the two-dimensional Navier--Stokes equation as firstly shown by \citet{Fox1971,Orszag1971}.

\section{Scale decomposition results of lead times other than $T_L$}\label{appA}
This section presents the scale-decomposition results of both models and visualisations for other lead times. The CC of the decomposed fields between the target and predictions by the cGAN and CNN for all lead times is listed in Table~\ref{table4}. The visualisations for $T=0.25T_L, 0.5T_L$, and $2T_L$ are shown in Figures~\ref{fig21a}(a), \ref{fig21a}(b), and \ref{fig21b}(c), respectively. For $T=0.25T_L$, as shown in Figure~\ref{fig21a}(a), there appears to be no significant difference in the performances of the cGAN and CNN for all scales. This is owing to the high correlation for the short lead time (CC = 0.8131 in Table~\ref{table4}), which enables the CNN model to generate accurate small-scale vorticity structures. However, the energy contained in the small-scale field is expected to be significantly underpredicted by the CNN from the enstrophy spectra in Figure~\ref{fig7}(a). For $T=0.5T_L$, the underprediction of the small-scale field by the CNN worsens, compared to the target and cGAN predictions. However, cGAN shows high accuracy both in Figures~\ref{fig21a}(a) and \ref{fig21a}(b) in the prediction of the small-scale structures unlike $T=T_L$ in Figure~\ref{fig9} of \secref{sub:pred1}. Finally, for $T=2T_L$ in Figure~\ref{fig21b}(c), the CNN fails to generate a proper prediction of the large-scale field, as well as the intermediate- and small-scale fields. Although the cGAN predictions were much better than those by CNN predictions, the overall predictions were not sufficiently good. However, in a statistical sense the cGAN appeared to generate reasonable predictions. 

\begin{figure}
	\centerline{\includegraphics[width=1.0\columnwidth]{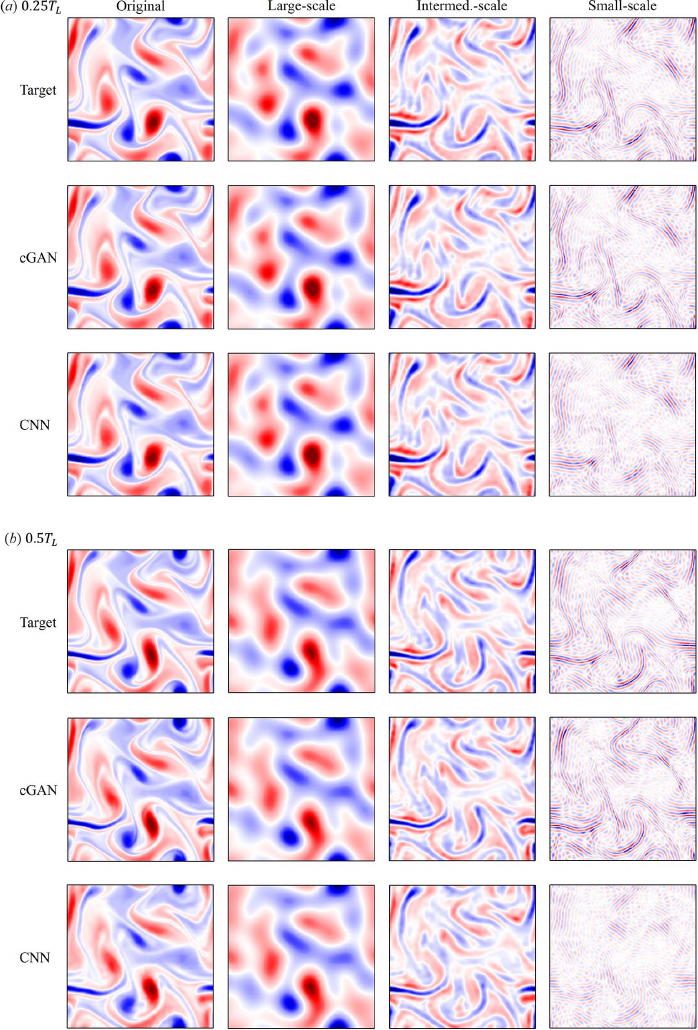}}
	\captionsetup{width=\columnwidth,justification=justified,labelsep=period}
	\caption{Scale decompositions of other values of $T$ (continued on the next page).}
	\label{fig21a}
\end{figure}

\begin{figure}
    \ContinuedFloat
	\centerline{\includegraphics[width=1.0\columnwidth]{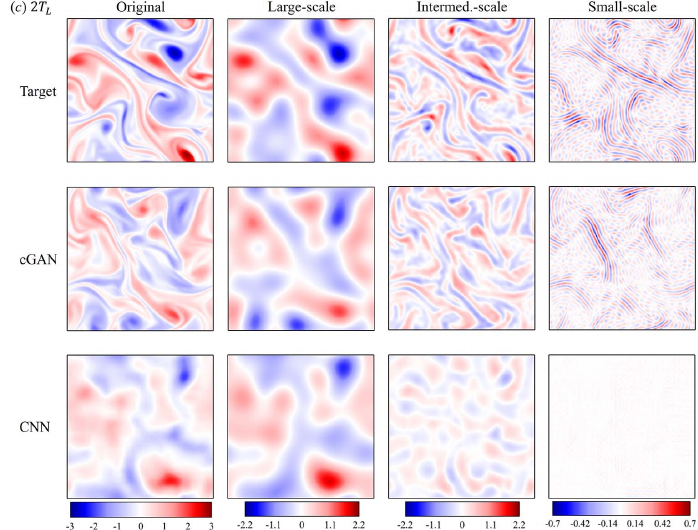}}
	\captionsetup{width=\columnwidth,justification=justified,labelsep=period,list=off,format=cont}
	\caption{Scale decompositions of other $T$.}
	\label{fig21b}
\end{figure}
\begin{table}\centering
	\begin{tabular}{c c c c c c c c c}
		lead time \;& \multicolumn{2}{c}{0.25$T_L$} \;& \multicolumn{2}{c}{0.5$T_L$} \;& \multicolumn{2}{c}{$T_L$} \;& \multicolumn{2}{c}{2$T_L$}\\
		\hline
		\;& cGAN \;& CNN \;& cGAN \;& CNN \;& cGAN \;& CNN \;& cGAN \;& CNN\\
		\hline
		Original \;& 0.9940 \;& 0.9898\;& 0.9708 \;& 0.9699 \;& 0.8547 \;& 0.8870 \;& 0.4530 \;& 0.5227\\[2pt]
		Large-scale \;& 0.9992 \;& 0.9983 \;& 0.9964 \;& 0.9952 \;& 0.9707 \;& 0.9750 \;& 0.6505 \;& 0.6542\\[2pt]
		Intermediate-scale \;& 0.9920 \;& 0.9869 \;& 0.9470 \;& 0.9449 \;& 0.6662 \;& 0.7166 \;& 0.0415 \;& 0.0292\\[2pt]
		Small-scale \;& 0.8956 \;& 0.8131 \;& 0.6348 \;& 0.5727 \;& 0.1148 \;& 0.0909 \;& -0.0008 \;& 0.0004\\
		\hline
	\end{tabular}
	\captionsetup{labelsep=period}
	\caption{Correlation coefficient between the target and prediction of the scale-decomposed fields depending on the lead time.}
	\label{table4}
\end{table}

\section{PredictionNet - Predictive accuracy on velocity statistics}\label{appB}
\begin{figure}
	\centerline{\includegraphics[width=1.0\columnwidth]{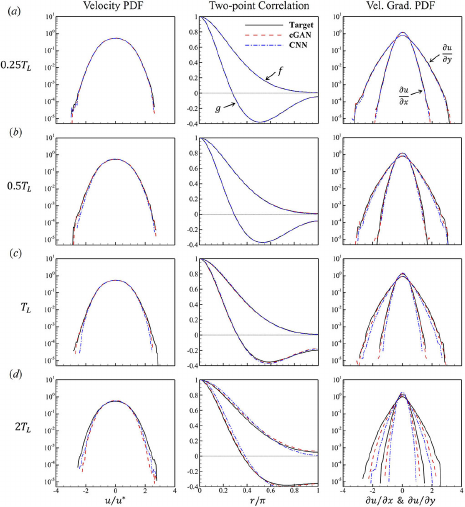}}
	\captionsetup{width=\columnwidth,justification=justified,labelsep=period}
	\caption{Comparison between the velocity statistics of the target and prediction results depending on $T$. (a) $T=0.25T_L$, (b) $T=0.5 T_L$, (c) $T=T_L$, and (d) $T=2T_L$ for the same input time point of $t_0$. The first, second, and third columns display velocity PDFs, longitudinal and transverse correlation functions, and velocity gradient PDFs, respectively.}
	\label{fig22}
\end{figure}

The performance of PredictionNet was investigated in the vorticity field, because 2D turbulence can be fully simulated using vorticity alone. In this section, the investigation of the performance of the developed network in predicting the velocity vectors is briefly discussed. In Figure~\ref{fig22}, the results of the velocity PDFs, longitudinal and transverse correlation functions ($f$ and $g$), and velocity gradient PDFs are compared between the cGAN and CNN for various lead times. The overall superiority of cGAN over the baseline CNN is observed again. In particular, for the velocity gradient, a performance difference similar to that of the vorticity was observed. Notably, neither the velocity PDF nor the two-point correlations show significant differences between the two models when compared to Figure~\ref{fig8} and the velocity gradient statistics. Evidently, the small-scale features in the velocity fields are less dominant than those in the vorticity fields; thus, the velocity field is easier to predict than the vorticity field. However, the cGAN model outperformed the CNN in terms of velocity prediction.

\section{Effect of adding an explicit loss to CNN}\label{appC}
In this section, we present the performance of a CNN model with an explicit loss reflecting the turbulence characteristics. Within the GAN scheme, the generator learns the system characteristics embedded in the data by minimizing an implicit loss based on the latent variable in the discriminator. This significantly enhanced PredictionNet compared to the baseline CNN model. However, even the CNN trained solely using MSE loss demonstrated reasonable prediction for relatively short lead times; the issue of poor performance across all lead times was in the small-scale structures. Hence, we investigate possibility of performance enhancement by explicitly enforcing statistics on the model, without introducing an additional network for adversarial learning. To achieve this, a spectrum loss term has been incorporated into the existing CNN, employing the following formula:
\begin{gather}\label{eqC1}
	L_{CNN}=\sigma_1 \mathbb{E}_{X\sim p(X)} \left[\|{Y^*- Y}\|_2^2\right]+\sigma_2 R(w_p)+\sigma_3 L_{spectrum}, \nonumber \\[3pt]
	L_{spectrum}=\mathbb{E}_{X\sim p(X)}\left[\|\log\left(\Omega_Y(k)\right)-\log\left(\Omega_{Y^*}(k)\right)\|_1\right],
\end{gather}
where $\Omega_Y(k)$ and $\Omega_{Y^*}(k)$ represent enstrophy spectra of the target and prediction, respectively. $\sigma_3$ modulates the strength of spectrum loss and two different values of $\sigma_3$,  $0.01/K$ (Spect 0.01) and $0.1/K$ (Spect 0.1), were tested, where the maximum valid wavenumber $K=64$.

\begin{figure}
	\centerline{\includegraphics[width=1.0\columnwidth]{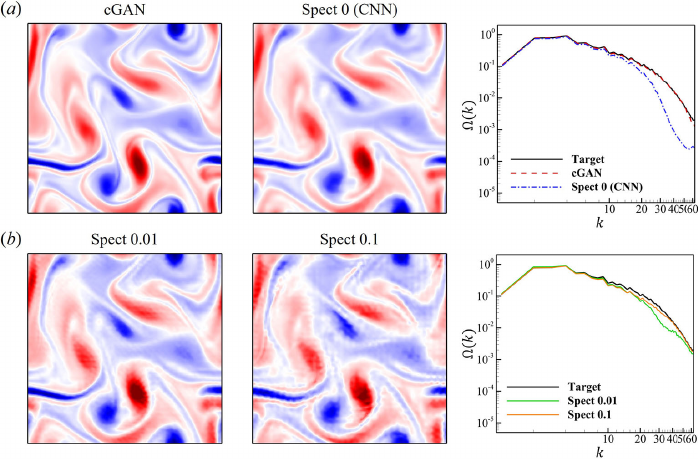}}
	\captionsetup{width=\columnwidth,labelsep=period}
	\caption{Visualised examples of prediction result and enstrophy spectra for $0.5T_L$ prediction using \\ (a) cGAN and Spect 0 (baseline CNN) and (b) spectrum loss added CNNs.}
	\label{fig23}
\end{figure}

In training models for $T=0.5T_L$, both cases exhibited convergence in the data loss. However, it's worth noting that during the training process, these cases displayed more fluctuations in the data loss compared to what was observed in Figure~\ref{fig6}. Figure~\ref{fig23} shows comparison of predicted fields and the corresponding enstrophy spectra that was explicitly given as a loss term in the objective function. In Figure~\ref{fig23}(a), for a direct comparison, results by cGAN and original CNN for $0.5T_L$ have been extracted from Figures~\ref{fig7} and \ref{fig8} in the main text. Figure~\ref{fig23}(b) displays the outcomes for the additional cases. Two noteworthy observations can be made out of these figures. Firstly, both Spect 0.01 and Spect 0.1 significantly improved the spectrum when compared to Spect 0. Secondly, despite the enhancement in spectrum, the visualizations show poor performance compared to Spect 0. When comparing other statistical metrics to those presented in Table~\ref{table2}, Spect 0.01 showed slightly inferior performance compared to Spect 0, with CC of 0.9644, MSE of 0.0477, and RMS of 0.9384. Spect 0.1 performed even worse, with CC of 0.9441, MSE of 0.0736, and RMS of 0.9179. It is important to highlight that although the RMS of Spect 0.01 seems closer to the target even than cGAN, it was caused by underpredictions in spectrum at intermediate and small scales and the slight overpredictions at large scales.
 
\begin{figure}
	\centerline{\includegraphics[width=0.42\columnwidth]{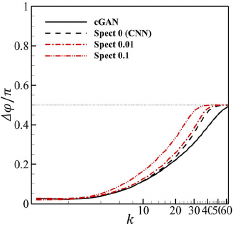}}
	\captionsetup{width=\columnwidth,labelsep=period}
	\caption{Prediction results of the phase error for $0.5T_L$ comparing cGAN, Spect 0,\\ and spectrum loss added CNNs.}
	\label{fig24}
\end{figure}

We observed that a consideration of an explicit spectrum loss in the training of CNN did not lead to better performance. We investigate the phase information, an essential information along with the spectrum. Figure~\ref{fig24} provides the phase error in which Spect 0.01 exhibits a slightly increased error compared to Spect 0, and Spect 0.1 reveals a significant degradation in the phase error. It can be inferred that explicitly introducing specific statistical characteristics of the system as loss terms in the model might lead to degradation in other statistics. One can potentially enhance performance by adding more explicit losses while conducting finer optimization. However, this approach requires an extensive parameter optimization. On the other hand, GANs, by implicitly learning system characteristics through competition with the discriminator, seem to effectively learn the statistical properties of turbulence. 

\section{Effect of the disturbance strength on the pre-trained surrogate model}\label{appD}
\begin{figure}
	\centerline{\includegraphics[width=1.0\columnwidth]{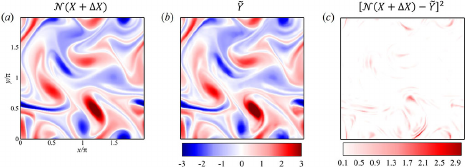}}
	\captionsetup{width=\columnwidth,justification=justified,labelsep=period}
	\caption{Example of visualising the effect of the disturbance strength on a pre-trained surrogate model using one test dataset with cGAN-$0.5T_L$ and $\Delta X_{rms}$=0.2$X_{rms}$. (a) Disturbance-added simulation ($\mathscr{N}(X+\Delta X)$), (b) disturbance-added prediction ($Pred(X+\Delta X)=\tilde{Y}$), and (c) squared difference of (a) and (b).}
	\label{fig25}
\end{figure}

For the training of the control network, PredictionNet (cGAN-$0.5T_L$), was used as a surrogate model. As mentioned in \secref{sub:control1}, if the strength of the disturbance $\Delta X$ is significantly large, then $X+\Delta X$ will deviate from the dynamics of the pre-trained PredictionNet, and the surrogate model will not work properly. Thus, the control effect cannot be properly evaluated as disturbance-added predictions, $Pred(X+\Delta X)=\tilde{Y}$, are vastly different from the results of the actual simulation, $\mathscr{N}(X+\Delta X)$, with $X+\Delta X$ as the initial condition. Based on the RMS of the input field, $\tilde{Y}$ was compared with $\mathscr{N}(X+\Delta X)$ using disturbances of various strengths ($\Delta X_{rms}$ = 0.1, 0.5, 1, 5, 10, and 20\% of $X_{rms}$). We could confirm that the surrogate model works properly for all testing cases. Therefore, an example using one of the test data is presented in Figure~\ref{fig25} for the 20\% case, which most likely deviates from the dynamics of PredictionNet. The two fields in Figures~\ref{fig25}(a) and \ref{fig25}(b) are qualitatively similar, and most of the area in Figure~\ref{fig25}(c), which shows the square of the difference between the two, is close to zero. Moreover, the values are not large even at positions where a relatively large error is observed. The MSE in Figure~\ref{fig25}(c) is 0.0582, which is reasonably small. In addition, there was little change in the structure of the disturbance field as the disturbance strength changed; there was only a difference in the degree of change in the field over time. Therefore, the disturbance-added solution for a specific target does not respond sensitively to the disturbance strength within the operating range of the surrogate model.

\bibliographystyle{jfm}
\bibliography{ref.bib}
	
\end{document}